\documentclass{article}

\usepackage{arxiv}

\usepackage[utf8]{inputenc} \usepackage[T1]{fontenc}    \PassOptionsToPackage{hyphens}{url}\usepackage[]{hyperref} \usepackage{url}            \usepackage{booktabs}       \usepackage{nicefrac}       \usepackage{microtype}      \usepackage{lipsum}		\usepackage{graphicx}
\usepackage[sort&compress,numbers]{natbib}
\usepackage{doi}
\usepackage{float}
\usepackage{multirow}
\usepackage[title]{appendix}
\usepackage{amsfonts}       \usepackage{amsmath}
\usepackage[inline]{enumitem}

\usepackage{indentfirst}

\usepackage{caption}
\DeclareCaptionType{equ}[][]

\usepackage{lineno}

\title{Esports Training, Periodization, and Software --\\a Scoping Review}

\date{\today}	

\author{ \href{https://orcid.org/0000-0003-3668-4638}{\includegraphics[scale=0.06]{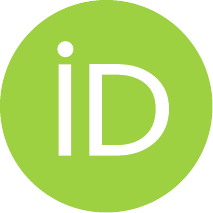}\hspace{1mm}Andrzej Białecki}\thanks{Corresponding Author} \\
	Warsaw University of Technology\\
	Warsaw, Poland \\
    \texttt{andrzej.bialecki.dokt@pw.edu.pl} \\
	\texttt{andrzej.bialecki94@gmail.com} \\
\And
	\href{https://orcid.org/0000-0002-4469-8004}{\includegraphics[scale=0.06]{orcid.pdf}\hspace{1mm}Bartłomiej Michalak} \\
	Józef Piłsudski University of Physical Education in Warsaw \\
	Warsaw, Poland \\
	\texttt{bartlomiej.michalak@awf.edu.pl} \\
	\And
	\href{https://orcid.org/0000-0002-2146-6198}{\includegraphics[scale=0.06]{orcid.pdf}\hspace{1mm}Jan Gajewski} \\
	Józef Piłsudski University of Physical Education in Warsaw \\
	Warsaw, Poland \\
	\texttt{jan.gajewski@awf.edu.pl} \\
}

\date{June 15, 2024}

\hypersetup{
pdftitle={Periodization Training and Tools in Esports a Literature Review},
pdfauthor={Andrzej Białecki, Bartłomiej Michalak, Jan Gajewski},
pdfkeywords={esports, sport science, periodization, training},
}

\providecommand{\parencite}[1]{}

\renewcommand{\parencite}[1]{\cite{#1}}

\begin{document}
\renewcommand{\figureautorefname}{Fig.}
\renewcommand{\subsectionautorefname}{Subsection}

\maketitle

\begin{abstract}

    Electronic sports (esports) and research on this emerging field are interdisciplinary in nature. By extension, it is essential to understand how to standardize and structure training with the help of existing tools developed by years of research in sports sciences and informatics. Our goal in this article was to verify if the current body of research contains substantial evidence of the training systems applied to training esports players.
    To verify the existing sources, we have applied a framework of scoping review to address the search from multiple scientific databases with further local processing.
    We conclude that the current research on esports dealt mainly with describing and modeling performance metrics spanned over multiple fragmented research areas (psychology, nutrition, informatics), and yet these building blocks were not assembled into an existing well-functioning theory of performance in esports by providing exercise regimes, and ways of periodization for esports.
\end{abstract}


\section{Introduction}
\label{sec:introduction}

Esports have proven to be a growing area of interest. This interest is not universally spread between varying disciplines. Gaming, esports, and interactive computer-simulated environments seem to be well established in the area of psychology \parencite{Banyai2019,Ramirez2020,Leis2020,Beres2023}. Sometimes, such psychological factors are related to physical characteristics of esports \parencite{Matesanz2023}. Finally, comprehensive reviews relate traditional sports to esports and verify biopsychosocial factors holistically \parencite{Shulze2023}. The same can be said for the broad Information and Communications Technologies area of research, where simulated environments help in robotics and AI research \parencite{Kaufmann2023}. Because of this, research on gaming and esports that deals with players at the height of their abilities is crucial. After all, in between areas of science, we can identify the algorithms, information systems, and data-based visualization tools that can support effective high-performance training in gaming and esports \parencite{Watson2021}.

However, some distinct parts of sports sciences are at the far end of this research's popularity. In traditional sports, topics such as training stages, periodization, stimuli/exercise selection, and control are well defined within the area of sports theory \cite{Sozanski2015,Kasper2019}. Let us assume that one would like to start strength training - there are multiple well-described methods for progression, rest times, and which exercises are best for this goal \parencite{RathiSharmaThapa2023,Thapa2023}. Similarly, researchers in sports define techniques that aim at improving regeneration through active and passive means \parencite{Adamczyk2023,Boguszewski2024}. Currently, in esports, it seems that there are no well-defined exercises, no definitions for intensity and load, and no way to test or verify the progression quickly and assess it against actual in-game performance.

While gaming and esports are areas where data is abundant and relatively easy to access by anyone with computer skills \parencite{URLBlizzardS2ClientProto,URLS2Prot2017,URLBoxcars2016,AWPYXeno2020}, we seem to know nothing about the internal workings of this area. Essential information on properly structuring esports training from start to finish remains to be discovered and made available to the broader community. There are limited readily available tools that assist in periodization and uncovering what kind of training structure is optimal for various game genres.

The ability to achieve the highest sports results is a derivative of a well-functioning and properly planned - and broadly understood - sports training system. According to the model assumptions, it consists of subsystems: forecasting, training, recovery and competition, qualification for sports, infrastructure and material support, and coaching staff. Therefore, it can be stated that only comprehensive solutions enable the achievement of sports mastery in every discipline \parencite{Sozanski2015}.

We perceive esports as a field that is expanding very fast, with many groups of interest and stakeholders, mainly players, media, tournament organizers, and others \parencite{Peng2020,Scholz2020,Scholz2019}. Games provide an everchanging environment that may be impossible to research effectively towards a comprehensive model of training at the highest level of play \parencite{Claypool2017,Wang2020,Zhong2022}. What if some of the skills that the players work for are instantly outdated by a new game release? In that case, we would like to emphasize the fundamental knowledge of exercises, scheduling, planning, and periodization that can be applied to any game.

Our goal for this work is to review the available literature in esports research, focusing on informatics (training tools, visualization, analytics, and feedback systems) and sports sciences (training, periodization, planning, and career stages). Works that have the potential for practical application in facilitating esports training are particularly interesting. \section{Methods}
\label{sec:methods}

Before performing any text-related analyses. To conform with the scoping review standards, we have manually collected data from scientific databases such as:
\begin{enumerate*}[label=(\arabic*)]
    \item ACM Digital Library,
    \item IEEE Xplore,
    \item Scopus,
    \item Springer Link,
    \item Web of Science.
\end{enumerate*}
For further processing, we leveraged the Python package ``litstudy'' offering a suite of tools for working with literature-based workflows \parencite{Litstudy2022}.

\subsection{Search Strategy}

In our efforts to perform a comprehensive search, we prepared specific inclusion criteria before searching the scientific databases. Based on our goal, we have decided to include all esports-related works published and available in scientific literature databases, focusing solely on works written in English. The search was constrained by the following query ``(esport OR esports OR e-sport OR e-sports)'' due to the varying naming used across the esports academic landscape. Later, all of these records would be filtered based on the inclusion criteria to fit the scope of our research goals. We set the search cutoff date at 2024.06.15.

\subsection{Data Processing and Inclusion Criteria}

Due to the overhead of defining complex queries in varying scientific search engines, we performed the rest of the analyses locally by leveraging the Python programming language and custom processing scripts with the ``litstudy'' package \parencite{Litstudy2022}. We ran automatic deduplication on the entire data set from multiple databases to identify the articles for further processing. Further processing warranted removing all of the works without the authors, and we have found that such works were most often article collections. Additionally, we have filtered out all articles that were not in English or contained ``l'esport'' in the title, abstract, or keywords.

Similarly, to ensure that all of the documents considered research on esports, the following keyword search was conducted: esport, esports, e-sport, e-sports, electronic sports, e-sports, pro-gaming, and professional gaming. Given the interdisciplinary area of esports research, our inclusion criteria were split between two interest groups. One dealing with informatics, and the other with the area of sports sciences. Two groups of keywords were defined: (1) tool, data analysis, data science, dataset, machine learning, artificial intelligence, visualization; (2) training, performance, cognitive training, neurofeedback, periodization, plan, and sport science. One caveat was that articles with the word ``training'' could have been found in both groups due to the word's broad meaning and its use in the context of ``neural network training''.

\begin{figure}[H]
    \centering
    \includegraphics[height=0.85\textheight]{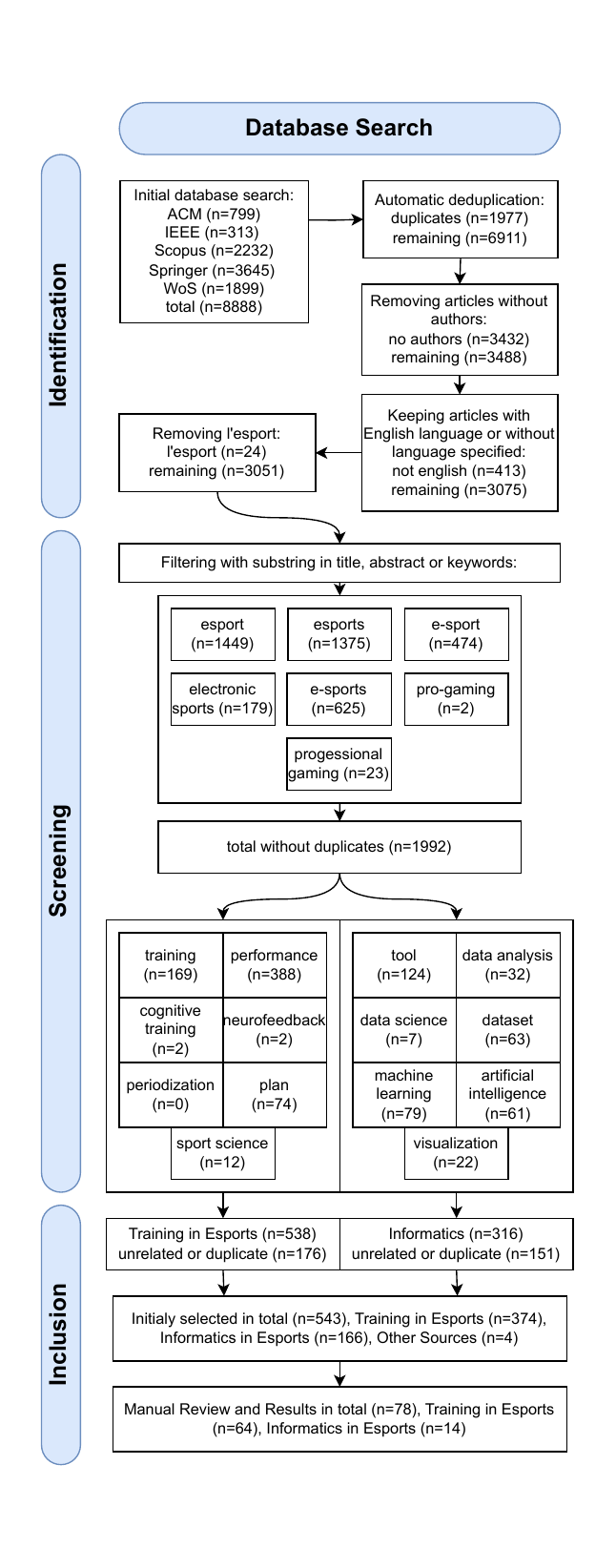}
    \caption{Scoping review article processing.}
    \label{fig:processing_pipeline}
\end{figure}

Finally, we merged both document sets with automatic duplicate removal. In the end, we reviewed all the articles left and manually selected the most applicable works to answer our research questions. The entirety of the process is visualized on \autoref{fig:processing_pipeline}.

\section{Results}

\begin{figure}[H]
    \centering
    \includegraphics[width=1\linewidth]{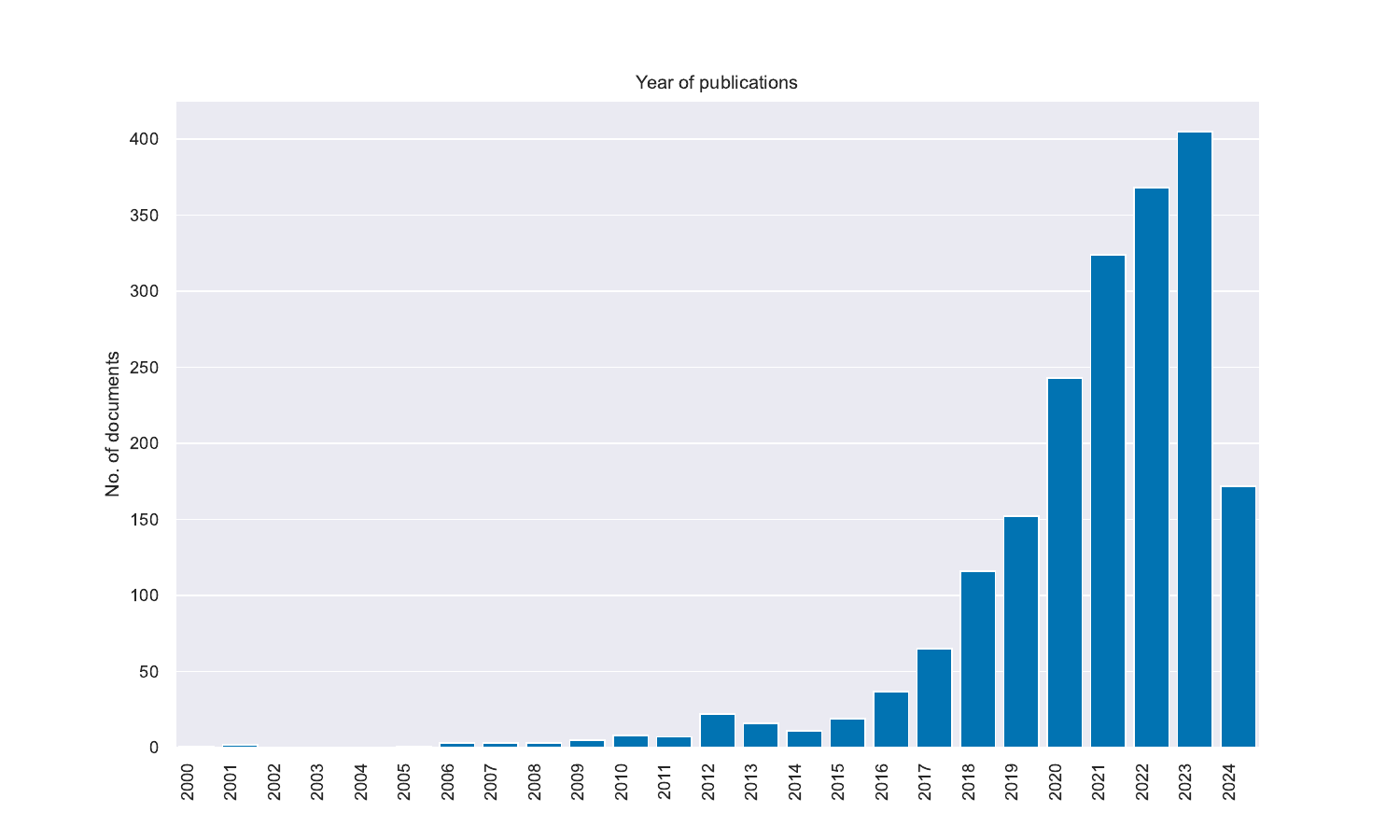}
    \caption{Year of publication histogram.}
    \label{fig:pub_histogram}
\end{figure}

\begin{figure}[H]
    \centering
    \includegraphics[width=0.8\linewidth]{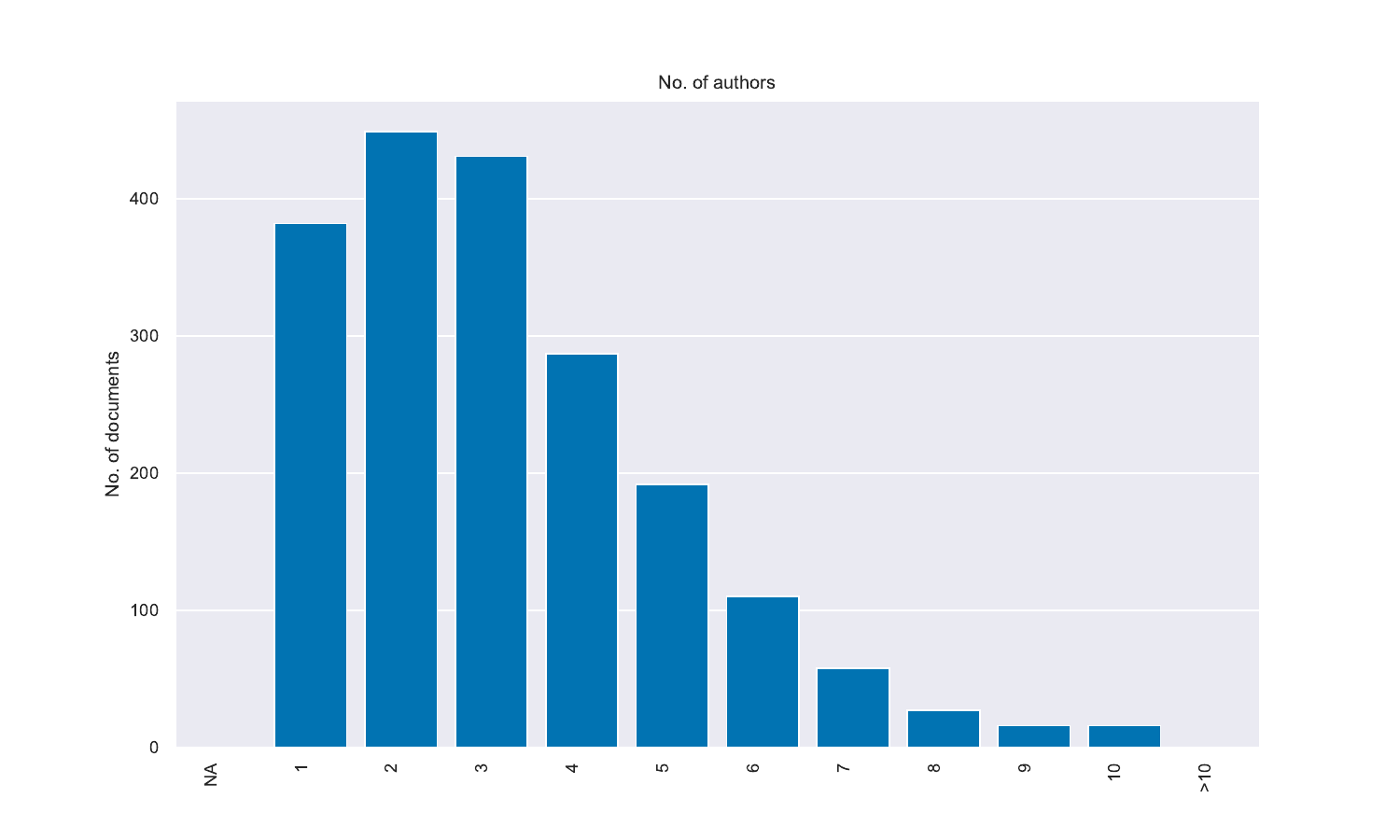}
    \caption{Number of authors histogram.}
    \label{fig:n_authors_histogram}
\end{figure}

While sources exist on esports curricula, their contents, and how to introduce esports into progressively more domains besides entertainment, more knowledge is still needed on how to structure esports training to facilitate the best possible individualized progress for esports players. Is it even possible to create a curriculum that could prepare coaches when there is no information on how to optimize training regimes and take care of upcoming esports athletes \parencite{Scott2022HigherED,Ibda2023SCHOOL,Fiskaali2020HighSchool}? Visual representation of the entire dataset is presented in \autoref{fig:pub_histogram}, and \autoref{fig:n_authors_histogram}.

\subsection{Training in Esports}

Some authors conducted work assessing how much time esports players devote to training \cite{Kari2016}. According to their work, the average training time of elite esports players per day is 5 hours and 28 minutes, of which just over 1 hour was related to typical physical training. Others reported far longer training times - as much as 9.34$\pm$1.12 hours a day \cite{Bayraktar2020}. Notably, the researchers set the criterion for selection into the test as requiring an esports license and representing their country in international competitions. There are also known cases where players train 12-14 hours a day, although some of this time might have been devoted to team meetings, video analysis, and strategic discussions \parencite{Kari2016}.

However, others showed that people who train more than 14 hours a week have significantly shorter visual reaction time and better accuracy than those who train below the established limit of 14 h/week \cite{Ersin2022}. Moreover, both indicators mentioned above were improved due to supplementation/intake of 3 mg/kg of caffeine, as confirmed by some studies \parencite{Sainz2020,Wu2024}. Improving reaction time is also possible by using appropriate training methods and tools. \cite{Qian2021} claim that reaction speed when using an effective prompt is significantly faster than when the prompt is invalid. Additionally, some authors drew attention to the need to use increasingly specialized training methods in esports \cite{Argiles2022}. Their topical review considered esports optometric factors such as screen time and digital eyestrain, visual skill demands, and perceptual-cognitive skills such as visual attention. The problem of a comprehensive approach to esports training can be seen in the work of \cite{Goulart2023}, who focused on nutrition, lifestyle, and cognitive performance in esports athletes. They demonstrated that esports athletes are consuming the recommended amount of protein, riboflavin, phosphorous, vitamin B12, and selenium performed significantly better over the 18 core NTx (cognition via 3-dimensional multiple objects tracking test (3DMOT) via Neurotracker X (NTx) software) sessions than those that did not meet the recommended amounts.

\subsubsection{Esports and Gaming, Mental Training}

Within all the publications that were qualified for this review, 38 referred to training aimed at developing the specialistic capabilities of esports players, i.e., those directly related to game achievements and performance. Among the selected works, the primary purpose of the research was to identify the length and frequency of training (playing games). Additionally, 12 works drew attention to the need to use imagery or mental training to improve the athlete's performance. Nine of twelve articles were unavailable for access without payment and were not reviewed further. Among the available articles, one dealt with brain-computer interface (BCI) controlled motor imagery training, confirming its positive impact on performance in experimental settings \parencite{Yakovlev2020}. In other cases, authors focus on explaining the Applied Model of imagery \parencite{Moritz2023}. Finally, as the authors explain how existing techniques like layered stimulus response training (LSRT) can be leveraged in the case of esports, they need to verify the technique's efficiency in practice \parencite{Cumming2023}.

\subsubsection{Physicality of Esports}

Works on physical training in preparing esports athletes constituted the second most extensive set of n = 18. This category also included works that focused on the quantitative and qualitative descriptions of players' physical activity and performance. Two of these 25 articles were only available with further payment and could not be manually reviewed. Among other studies, one piloted a 8-week intervention based on introducing physical exercise to the training routine of high-level esports players. Researchers found that physical activity reduced players' fatigue perception and improved their physical performance \parencite{SanzMatesanz2024}. Among adolescents, most of the respondents declared that they are physically active. However, it is unclear how many of them are active esports players and not recreational gamers \parencite{Szepne2019physical}.

Similarly, other authors investigated the attitudes of grassroots players towards physical activity \parencite{Fletcher2021}. Finally, other authors found that 6-minute active breaks in the session of practice improved executive functions in players \parencite{Difrancisco-Donoghue2021}. Generally, research seems to recognize the physical aspects of esports and describe them \parencite{Ekdahl2022}. Despite that, there is no clear evidence of leveraging physical training to improve esports performance.

On the other hand, other studies dive into the medical aspects of injury prevention and pain within competitive gamers and esports athletes \parencite{Pereira2019,Law2023224,McGee2021415}. Authors underline the importance of medical attention in the case of esports athletes to prevent early retirement and financial consequences \parencite{Migliore2021}. Research in this direction uncovered that players tend to train less when in the presence of pain \parencite{Lindberg2020Pain}.

Some authors found that short sprint exercises impacted cognitive performance in amateur esports players positively \parencite{Manci2024}. Studies performing systematic reviews found that despite multiple articles reporting on esports players' engagement in physical activity, their reported activity was only a small portion of the main training components \parencite{Lam2020Eathletes}. In other cases, players were described as a low-active group, while high-level professionals were found to have higher levels of physical activity \parencite{Voisin202232}. The vast majority of studies focus on self-reported data for physical activity levels. While this paints a picture of physically active esports athletes, such methods do not provide an entirely objective view of the situation.

On the other hand, some works considered this limitation and outfitted their respondents with accelerometer devices to asses their physical activity levels. They found that participants overstate their physical activity \parencite{Nicholson2024}. It seems that most of the players engage in physical activity as a way to stay healthy rather than to improve their performance \parencite{Kari2016}. Finally, physical activity is recognized as a potentially fruitful avenue of research for esports \parencite{Ketelhut2021}.

\subsubsection{Various Factors of Esports Performance}

Twelve publications concerned the impact of reaction time on gameplay results or broadly understood game performance. The analyzed works examined the impact of various methods and techniques for improving reaction time and the impact of equipment, e.g., mouse weight. Twelve publications concerned nutrition and supplementation of various products and nutrients. Most articles assessed what and how high-performance players eat, including supplementation of selected ingredients, e.g., caffeine, which affects their performance during competition. It is also worth emphasizing that in individual publications, it was possible to recognize opinions about an insufficient level of knowledge about nutrition in esports \parencite{Trocchio2021}.

Eighteen works concerned health and mental health. The analyzed works usually presented opinions on the perception of one's health condition on the results achieved and an analysis of esports players' opinions on whether mental and physical well-being affects the improvement of gaming results. Moreover, the published works included information on virtual athletes' healthy lifestyle habits. Sleep is also essential for health, and its impact on performance has been described in 10 studies. Moreover, these articles tried to characterize the best players' sleep length and quality of sleep.

Unfortunately, no publications regarding the periodization of training or esports were found. However, one work comprehensively described the training structure and presented a performance model \parencite{Nagorsky2020}. Some research suggests a strong need to create detailed structures and training methods based on real sports achievements \cite{Bialecki2022}. A similar position was expressed by more authors, who uncovered that organized training and its structure could influence the performance of esports athletes but also improve their health \cite{Bikas2023}.

\subsection{Informatics in Esports}

Esports change constantly due to the software patches, in-game changes, and advances to the technology used in game design \parencite{He2021Updates,Yu2021Updates}. It is clear that esports, in their essence, are a fully automated sport; each of the peripheral devices acts as an interface for the computer, where the rules of competition are implemented within the simulated environment. Due to this, access to data is abundant. There are multiple available datasets spanning game genres and game titles. Some focus on tournament games \parencite{Białecki2023SC2EGSetDataset}, or ladder games \parencite{Wu2023Dataset} in StarCraft 2. Others work with games that could be considered board games and computer games \parencite{Gao2022Dataset}. Some publish video data from League of Legends \parencite{Tanaka2021Dataset,Xu2023Dataset}, while others deal with Counter-Strike \parencite{Xenopoulos2022dataset}. There are undoubtedly many of them. These datasets are part of the vital ecosystem of infrastructure that is needed to develop the field of esports informatics. Creating automated systems with capabilities to reliably provide immediate feedback to the players, coaches, or other stakeholders could assist lower-tier players in their development.

Moreover, multiple works define data visualization experiences for gaming and esports. These tools provide technology that helps to see the often nuanced in-game environments in an easily digestible way \parencite{Horst2021,Afonso201933069}. Post-game reviews can be crucial in understanding the pain points in performance; leveraging this information by experienced players and coaches should be the ultimate goal \parencite{Kim202134189}. Despite that, such implementations are also applicable to other uses, e.g., betting \parencite{Spiricheva2019}, or spectatorship \parencite{Broek2019}. In the end, research exists on leveraging the varying algorithms to assist, defeat the human players, or to play against other agents/bots \parencite{Font2019285}.

\section{Discussion}
\label{sec:discussion}

\subsection{From Sports to Esports}

We raise the question of whether transferring the already established structures and systemic solutions to electronic sports is possible. If so, should all of these elements be copied and transferred to virtual competition? Our primary task is to focus on finding an appropriate system for selection and recruitment, a well-planned multi-year training process with a programmed duration for each training stage, appropriately distributed periodization of smaller training cycles, the choice and application of the most suitable methods, forms, and means of training, the proper proportion between training and rest, and the development of the psychomotor and cognitive abilities of future esports champions. With the help of IT tools, it is also possible to introduce an appropriate system for training control and analyzing tactical actions during sports competitions. This would contribute to creating a model of a champion in esports.

\begin{figure}[H]
    \centering
    \includegraphics[width=0.6\linewidth]{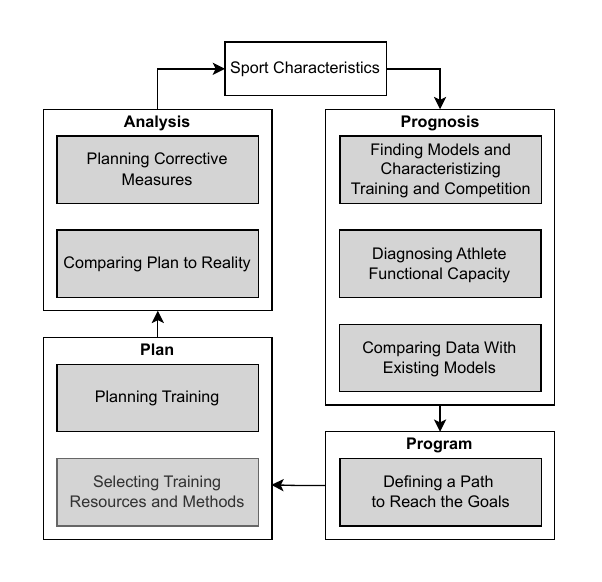}
    \caption{Diagnosis prognosis plan, as adapted from \parencite{Sozanski2015}.}
    \label{fig:diagnosis_program}
\end{figure}

Coaching staff highly seeks iterative processes of improvement in sports. Out of the 374 articles related to training in esports, 64 reports were considered the most related to the topic. However, juxtaposing the content of these publications with the ``diagnosis, program, plan'' model as adapted from \cite{Sozanski2015}, a very selective approach was noticed, focused mainly on single parameters or factors that are not a reference to a complete system of training in esports. This iterative system is presented in \autoref{fig:diagnosis_program}. This suggests to us that the current state of knowledge on training in esports is underdeveloped. Such a knowledge gap leaves the immense esports industry without the fundamental understanding required to develop future champions.

\subsubsection{Training Stimuli in Esports}

In widely known traditional sports, competitive goals determine the training process. This goal is often to achieve the best possible performance or tournament placement. Reaching mastery in any sport requires substantial training time comprised of a wide array of controlled stimuli. Psychophysiological adaptations accompany the effective use of these tools; the so-called ``form'' comes as a result, and the outcome should be a high competition placement. Due to the complexity of these processes, working with sports requires the right approach and time for these adaptations to occur. Sports sciences and theoretical approaches guide the optimal time to begin training in various sports.

\subsubsection{When to Start Training?}

First, there are three types of traditional sports: early, normal, and late. Involvement in early sports often starts at just 3-4 years old for upcoming athletes. In this category, the highest performance is reached at 15-17 years of age, e.g., gymnastics, swimming, tennis, and others. Sports assigned to the ``normal'' category start at 8-10 years old. These are, for example, track and field, football, volleyball, and others, where the highest performance comes at 18-21 years of age. Finally, in late sports, the start of training is marked at 13-14 years old, and the highest performance comes after 22 years of age, e.g., combat sports, weightlifting, long-distance running, and others \parencite{Goodway2015}.

Secondly, the right time to start training is bound to the anthropologically set level of development/maturity. Molding of physical abilities and especially coordination skills may be subject to ``critical times'', i.e. intervals at which it is best to apply certain stimuli to achieve maximal adaptation. Not leveraging this timing can trump the ability of an athlete to adapt further and reach their full potential \parencite{Szymanska2016}. These natural capabilities act as a base for formulating and defining stages in sports training; by extension, these stages include recommendations for methods, forms, and training means. Standardized stages of sports training and development emphasize the educational and fostering character of psychophysical capabilities through play. In this regard, it is essential to remember about other areas of life and other sports, as overspecializing in one area may not be beneficial. They primarily consider that in the process of anthropological development, the final sporting activity may not have been selected.

In their studies on esports training and physical exercise, other authors obtained data from a sample of 115 professionals and high-level esports players, with an average age of 20.8 years \cite{Kari2016}. Motivation in amateur esports players was studied by \cite{Arslan2024Motivation} - the average age of people participating in virtual competition was also similar and amounted to 20.15 years of age. When looking at amateur players who do not necessarily compete in esports, 24 fans at the Intel Extreme Masters (IEM) 2018 esports world championship in Katowice asked about their experience. It indicated that the average age of experience in gaming is, on average, ten years. The average age of participants was 19 years old. This indicates the average experience to be nine years. Participants also declared their first competition at about 14 years of age \parencite{Jasny2019}.

\subsubsection{Approaching Long Term Development}

\begin{figure}[H]
    \centering
    \includegraphics[width=0.6\linewidth]{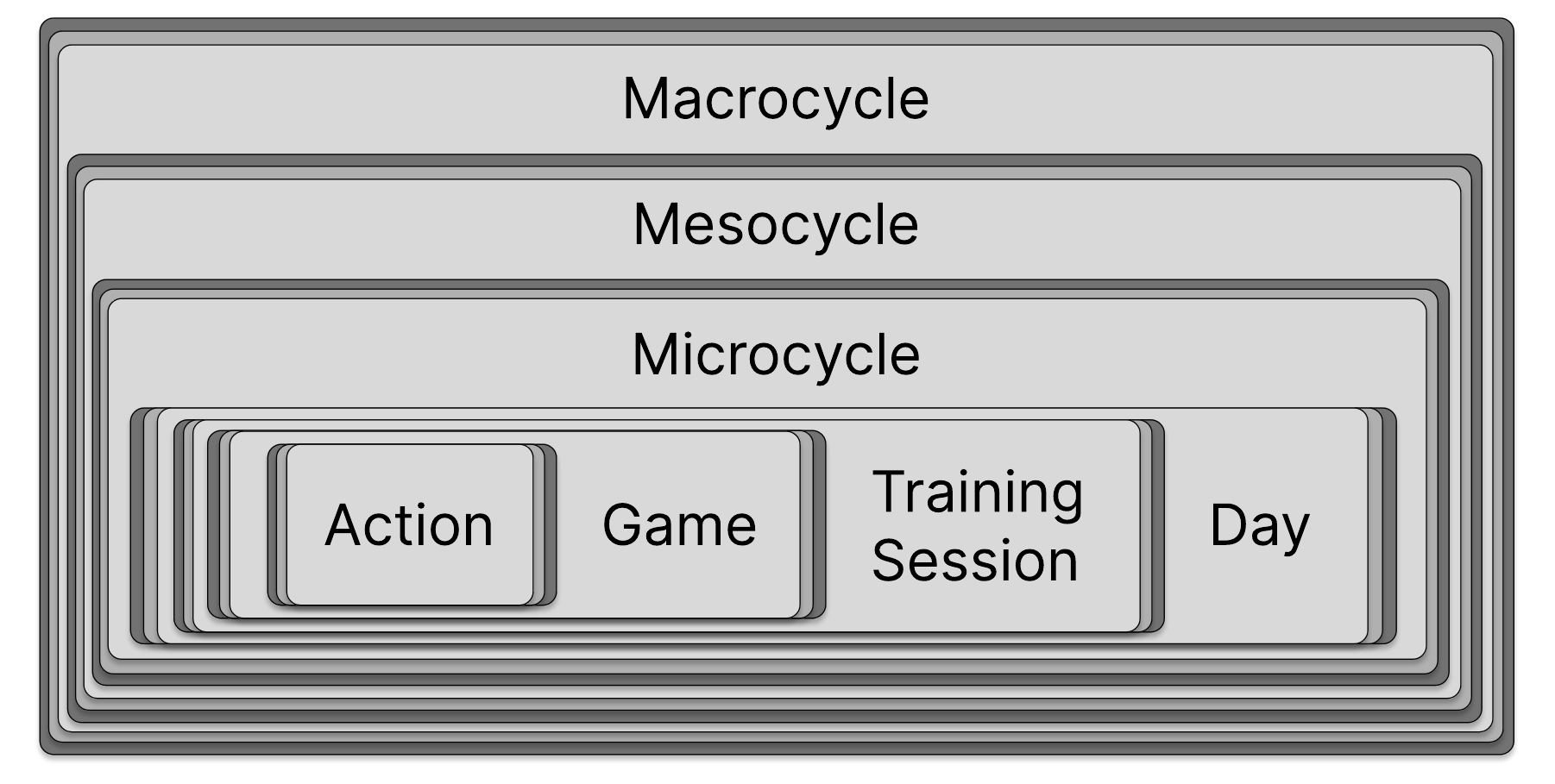}
    \caption{Compositional structure of sports.}
    \label{fig:compositional_structure_sports}
\end{figure}

Thirdly, selecting the optimal training load required to reach mastery is essential. According to \cite{Anders2016PEAK}, it takes roughly 10 000 hours and at least ten years of training to be an expert in a given activity. Of course, this does not precisely reflect reality, as everything depends on the athlete's selected sport and other individual features \parencite{Brenner2016}. In sports theory, when viewing training as a process, stimuli are evaluated and planned within a compositional structure as shown in \autoref{fig:compositional_structure_sports}.

Current concepts regarding the development of adolescent athletes are aimed at long-term athletic development (LTAD). They show multiple advantages while minimizing the possibility of premature career ending \parencite{Myer2016,Myer2015}. The LTAD model consists of solutions for anyone, not only the elite athletes but also people who engage in sporting activities as a base for their fitness and wellness \parencite{Granacher2017}. This model is built around seven stages of sports training:
\begin{enumerate*}[label=(\arabic*)]
    \item Active start,
    \item FUNdamental,
    \item Learn to train,
    \item Training to Train,
    \item Training to Compete,
    \item Training to Win,
    \item Retirement/Retraining \parencite{Brenner2016,Balyi2001,BalyiLTAD2013}.
\end{enumerate*}
In 2014, the United States Olympic Committee, along with their national administration, used the rules of LTAD to create the American Model of Development. In their work, they defined five stages that consist of:
\begin{enumerate*}[label=(\arabic*)]
    \item Discover, Learn, and Play (ages 0-12),
    \item Develop and Challenge (ages 10-16 years),
    \item Train and Compete (ages 13-19 years),
    \item Excel for High Performance or Participate and Succeed (ages $\ge$15 years),
    \item Mentor and Thrive for Life \parencite{Brenner2016}.
\end{enumerate*}

\subsubsection{Stimuli Effectiveness and Selection in Sports}

Traditional sports have a system of selection that is often ruthless for children who develop late and do not provide immediate results for their clubs or teams. Looking at computer games and esports from the perspective of sports science and traditional sports could be beneficial. Playing into the well-defined stages of human development and ``critical times'', beginning esports training as soon as possible should be researched. This is primarily due to cognitive abilities and movement coordination development between 6-12 years of age. Similarly, it is said that socializing at age 11-19 is essential, and long hours of sedentary training could act contrary to this. We recommend intensive research into how existing LTAD and other recommendations could be applied to successful esports development \parencite{Meredith2015,Fuhrmann2015}.

\subsection{From Esports to Sports}

The new technologies, including simulated environments, virtual reality, and extended or augmented reality, do not function in a vacuum. Where there is a new technology, people innovate. Such technologies hold high potential in providing training tools for traditional sports, rehabilitation, and many more \parencite{Sawan2020MRAR}. Creating specific training tools is possible, for example, designing a bow device to facilitate a virtual training of archery \parencite{Masasuke2023ARCHERY,Masasuke2022ARCHERY}.

Many advancements exist in the area of reinforcement learning. These systems are capable of beating human opponents in drone racing \parencite{Kaufmann2023}, Go \parencite{Silver2016}, chess and shogi \parencite{Schrittwieser2020,Silver2018}, StarCraft 2 \cite{Vinyals2019350}, Dota 2 \parencite{openai2019dota2largescale,raiman2019longtermplanning}. What if they were used to provide fast and actionable feedback to the human players? What are the limits of human performance with such assisting tools? Why not leverage these agents to detect and recommend action to enhance the human experience and plan training and educational efforts? Transitioning into the realm of optimizing an artificial intelligence model to play games and defining a task to optimize the human player's performance.

This kind of model could consider not only the immediate and short-term actions but additionally the long-term development of the player and many other players in the ecosystem. Aligning such systems with the existing knowledge in the theory of sports would mean creating a model that can assist with:
\begin{enumerate*}[label=(\arabic*)]
    \item action by action, decision by decision feedback,
    \item game-by-game feedback on tactics, strategies, and performance,
    \item session by session feedback,
    \item day-by-day feedback,
    \item microcycle (weekly) feedback,
    \item mesocycle (monthly) feedback,
    \item macrocycle (yearly) feedback, and more!
\end{enumerate*}

\begin{figure}[H]
    \centering
    \includegraphics[width=0.9\linewidth]{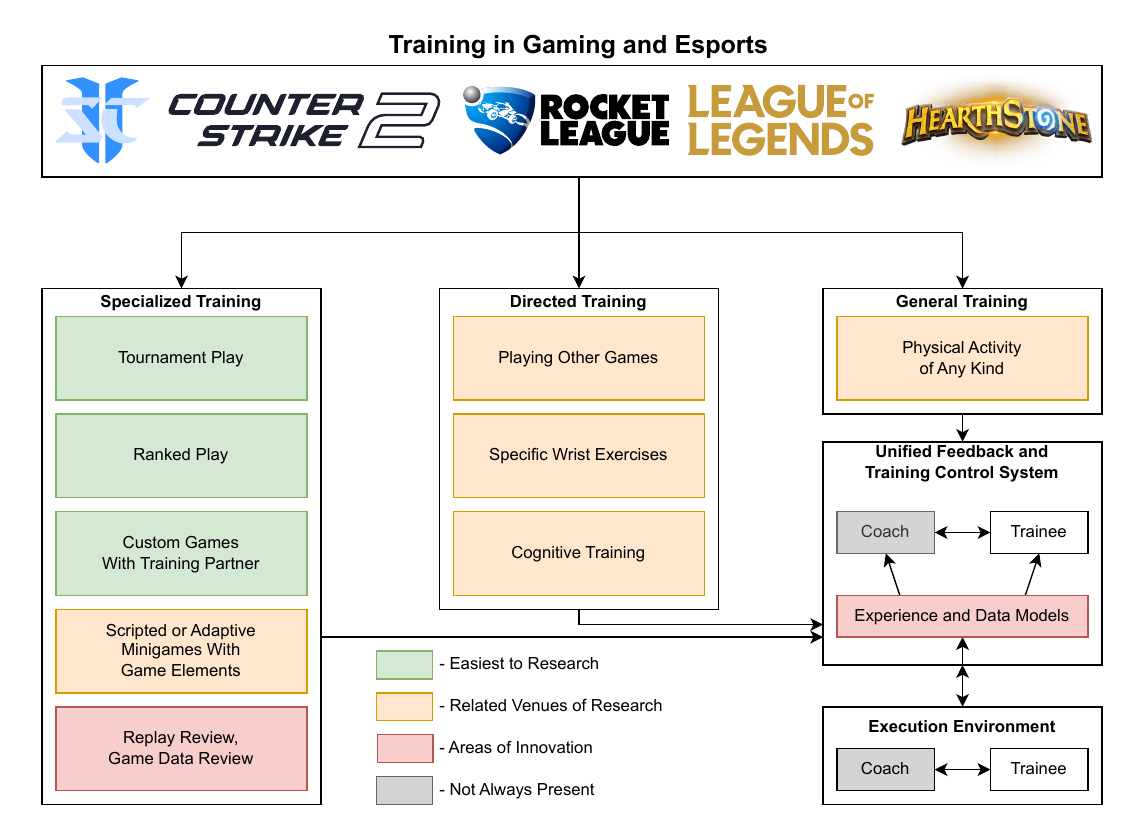}
    \caption{Model system of training in esports, partially adapted from \parencite{Sozanski2015}. Game logos attached for illustrative purposes.}
    \label{fig:perfect_training_system}
\end{figure}

\begin{figure}[H]
    \centering
    \includegraphics[width=0.65\linewidth]{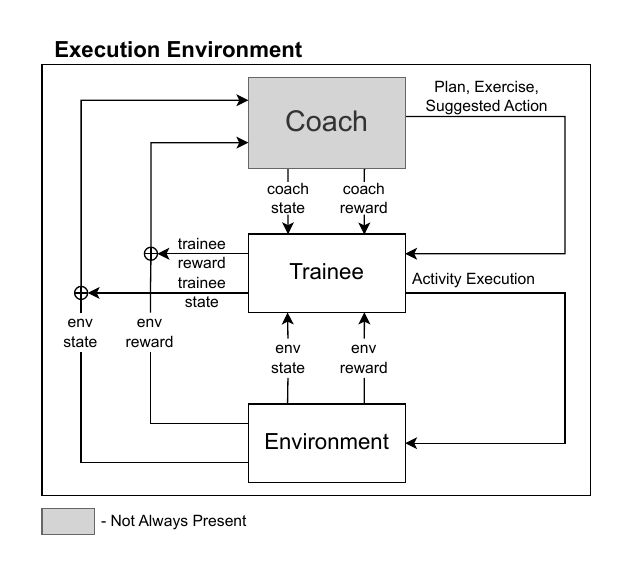}
    \caption{Model of an execution environment. Adapted from popular reinforcement learning explanatory diagrams.}
    \label{fig:execution_environment}
\end{figure}

Moreover, systems like that should have the capability to match the closest and related exercises to alleviate any issues in the players' performance effectively. Another diagram assisting with the understanding of this proposed system is shown in \autoref{fig:perfect_training_system}. Interestingly enough, the execution environment, as seen in \autoref{fig:execution_environment}, adapts the ideas from reinforcement learning (RL). We view it as a middle-ground between the most basic model diagram of RL as described by \parencite{Sutton2018} and how sports operate in reality.

Solutions of this kind are generally described as dynamic difficulty adjustment (DDA). Some researchers found that dynamic DDA contributed to heightened engagement \parencite{Damastuti2024}. Similarly, when looking at platform games, researchers found that applying performance-based DDA and artificial neural network (ANN) DDA allowed players to achieve a more enjoyable in-game experience. Additionally, authors recommend integrating physiological sensors for further research \parencite{Rosa2023}. In the end, DDA research discussed here focuses primarily on heightening the player enjoyment rather than optimizing the player performance \parencite{Silva2017}. It does not seem that researchers are focusing on how to apply such systems in a long-term development of a human player with a complete overview of various metrics longitudinally. Similarly to the conclusions made by authors of a study performing a systematic review on DDA \parencite{Mortazavi2024}, we urge researchers to attempt to apply DDA systems focusing on the structure of sports training as seen in \autoref{fig:compositional_structure_sports}.

\section{Limitations}
\label{sec:limitations}

Despite our study being a comprehensive literature review, it has limitations. We acknowledge that we may not have used other scientific literature databases in our \nameref{sec:methods}. Due to the dynamic nature of esports, it is unclear if information on practical and effective training is not already present. Gaming and esports communities tend to share their experiences in online tutorials and pass them down through video streaming or peer-to-peer learning within smaller esports communities. We recognize that while we did our best to look for all of the materials concerning esports. It may be impossible to collect all of them at any given time due to the nature of the academic indexing. We might have missed some articles concerning esports and player development.

To sum up, the cutoff point for the materials under investigation was placed at 2024.06.15. Some of the more recent publications might not have been included in our review. This is apparent and visible on \autoref{fig:pub_histogram}. Additionally, due to the esports interdisciplinary nature, some categories of articles may have overlapped. In other cases, articles that did not provide sufficient indexing information may have been ommited.
\section{Conclusions}

We conclude that the literature on data analyses, processing, and visualizations are extensive. These topics are a part of the broader subsystem of training control. Despite that, there is limited information, and the literature on periodization, planning training, and specialized exercises is insufficient.

There is a need to start researching the periodization of the training process in esports. The incomplete structure and inaccurate training characteristics in esports pose a considerable challenge for coaches and players. Therefore, organizing knowledge and filling gaps in broadly understood esports training is extremely important, following the example of well-established and existing knowledge in traditional sports. We urge researchers from varying backgrounds to join forces to put all of the building blocks of a successful training system together and start working on validating specific exercises and training methodologies.

Unfortunately, due to the complex nature of esports encompassing all possible games, defining a standardized approach that would fit all of them may not be possible. We encourage researchers to recognize these differences and start validating exercises that could fit into the existing periodization frameworks by defining them as general, directed, or specialized in the context of their selected games. Additionally, some games are discontinued, and technology depreciates; some training methodologies may not transfer to other games or game genres. Finding a way to conduct training in esports to create flourishing and long-lasting opportunities will be necessary.

Finally, it is tough to discern between studies that relate their findings to recreational gamers, competitive gamers, and esports athletes. Competitive play is not equal to esports performance. We urge researchers to clearly define their participant groups and the context of their research to make it easier to compare and contrast findings across studies, especially since all of the games have the potential to be played at a recreational level, competitive level, or professional level.

\section*{Declarations}

\subsection*{Funding}
No external funding was received for this study.

\subsection*{Conflicts of interest/Competing interests}
Authors declare no conflict of interest.

\subsection*{Authors' contributions}

\begin{itemize}
    \item Conceptualization, Andrzej Białecki;
    \item Methodology, Andrzej Białecki, Bartłomiej Michalak and Jan Gajewski;
    \item Software, Andrzej Białecki;
    \item Validation, Andrzej Białecki and Jan Gajewski;
    \item Formal analysis, Andrzej Białecki and Bartłomiej Michalak;
    \item Investigation, Andrzej Białecki and Bartłomiej Michalak;
    \item Resources, Andrzej Białecki;
    \item Data curation, Andrzej Białecki;
    \item Writing---original draft preparation, Andrzej Białecki and Bartłomiej Michalak;
    \item Writing---review and editing, Andrzej Białecki and Bartłomiej Michalak;
    \item Visualization, Andrzej Białecki;
    \item Supervision, Andrzej Białecki and Jan Gajewski;
    \item Project administration, Jan Gajewski;
    \item Funding acquisition, Jan Gajewski
\end{itemize}

\bibliographystyle{IEEEtran}
\bibliography{IEEEabrv,sources.bib}

\begin{thebibliography}{10}
\providecommand{\url}[1]{#1}
\csname url@samestyle\endcsname
\providecommand{\newblock}{\relax}
\providecommand{\bibinfo}[2]{#2}
\providecommand{\BIBentrySTDinterwordspacing}{\spaceskip=0pt\relax}
\providecommand{\BIBentryALTinterwordstretchfactor}{4}
\providecommand{\BIBentryALTinterwordspacing}{\spaceskip=\fontdimen2\font plus
\BIBentryALTinterwordstretchfactor\fontdimen3\font minus
  \fontdimen4\font\relax}
\providecommand{\BIBforeignlanguage}[2]{{%
\expandafter\ifx\csname l@#1\endcsname\relax
\typeout{** WARNING: IEEEtran.bst: No hyphenation pattern has been}%
\typeout{** loaded for the language `#1'. Using the pattern for}%
\typeout{** the default language instead.}%
\else
\language=\csname l@#1\endcsname
\fi
#2}}
\providecommand{\BIBdecl}{\relax}
\BIBdecl

\bibitem{Banyai2019}
\BIBentryALTinterwordspacing
F.~B{\'a}nyai, M.~D. Griffiths, O.~Kir{\'a}ly, and Z.~Demetrovics, ``{The
  Psychology of Esports: A Systematic Literature Review},'' \emph{Journal of
  Gambling Studies}, vol.~35, no.~2, pp. 351--365, 06 2019. [Online].
  Available: \url{https://doi.org/10.1007/s10899-018-9763-1}
\BIBentrySTDinterwordspacing

\bibitem{Ramirez2020}
\BIBentryALTinterwordspacing
M.~R. Ismael Pedraza-Ramirez, Lisa~Musculus and S.~Laborde, ``{Setting the
  scientific stage for esports psychology: a systematic review},''
  \emph{International Review of Sport and Exercise Psychology}, vol.~13, no.~1,
  pp. 319--352, 2020. [Online]. Available:
  \url{https://doi.org/10.1080/1750984X.2020.1723122}
\BIBentrySTDinterwordspacing

\bibitem{Leis2020}
\BIBentryALTinterwordspacing
O.~Leis and F.~Lautenbach, ``{Psychological and physiological stress in
  non-competitive and competitive esports settings: A systematic review},''
  \emph{Psychology of Sport and Exercise}, vol.~51, p. 101738, 2020. [Online].
  Available: \url{https://doi.org/10.1016/j.psychsport.2020.101738}
\BIBentrySTDinterwordspacing

\bibitem{Beres2023}
\BIBentryALTinterwordspacing
N.~A. Beres, M.~Klarkowski, and R.~L. Mandryk, ``{Playing with Emotions: A
  Systematic Review Examining Emotions and Emotion Regulation in Esports
  Performance},'' \emph{Proceedings of the ACM on Human-Computer Interaction},
  vol.~7, no. CHI PLAY, 10 2023. [Online]. Available:
  \url{https://doi.org/10.1145/3611041}
\BIBentrySTDinterwordspacing

\bibitem{Matesanz2023}
\BIBentryALTinterwordspacing
M.~Sanz-Matesanz, G.~M. Gea-García, and L.~M. Martínez-Aranda, ``{Physical
  and psychological factors related to player's health and performance in
  esports: A scoping review},'' \emph{Computers in Human Behavior}, vol. 143,
  p. 107698, 2023. [Online]. Available:
  \url{https://doi.org/10.1016/j.chb.2023.107698}
\BIBentrySTDinterwordspacing

\bibitem{Shulze2023}
\BIBentryALTinterwordspacing
M.~M. Jeffrey~Shulze and O.~Ruvalcaba, ``{The Biopsychosocial Factors That
  Impact eSports Players' Well-Being: A Systematic Review},'' \emph{Journal of
  Global Sport Management}, vol.~8, no.~2, pp. 478--502, 2023. [Online].
  Available: \url{https://doi.org/10.1080/24704067.2021.1991828}
\BIBentrySTDinterwordspacing

\bibitem{Kaufmann2023}
\BIBentryALTinterwordspacing
E.~Kaufmann, L.~Bauersfeld, A.~Loquercio, M.~M{\"u}ller, V.~Koltun, and
  D.~Scaramuzza, ``{Champion-level drone racing using deep reinforcement
  learning},'' \emph{Nature}, vol. 620, no. 7976, pp. 982--987, 08 2023.
  [Online]. Available: \url{https://doi.org/10.1038/s41586-023-06419-4}
\BIBentrySTDinterwordspacing

\bibitem{Watson2021}
\BIBentryALTinterwordspacing
B.~Watson, J.~Spjut, J.~Kim, J.~Listman, S.~Kim, R.~Wimmer, D.~Putrino, and
  B.~Lee, ``{Esports and High Performance HCI},'' in \emph{Extended Abstracts
  of the 2021 CHI Conference on Human Factors in Computing Systems}, ser. CHI
  EA '21.\hskip 1em plus 0.5em minus 0.4em\relax New York, NY, USA: Association
  for Computing Machinery, 2021. [Online]. Available:
  \url{https://doi.org/10.1145/3411763.3441313}
\BIBentrySTDinterwordspacing

\bibitem{Sozanski2015}
H.~Sozański, J.~Sadowski, and J.~Czerwiński, \emph{{Podstawy Teorii i
  Technologii Treningu Sportowego}}.\hskip 1em plus 0.5em minus 0.4em\relax
  Akademia Wychowania Fizycznego Józefa Piłsudskiego Filia w Białej
  Podlaskiej, 2015, vol.~2.

\bibitem{Kasper2019}
\BIBentryALTinterwordspacing
K.~Kasper, ``{Sports Training Principles},'' \emph{Current Sports Medicine
  Reports}, vol.~18, no.~4, 2019. [Online]. Available:
  \url{https://doi.org/10.1249/jsr.0000000000000576}
\BIBentrySTDinterwordspacing

\bibitem{RathiSharmaThapa2023}
\BIBentryALTinterwordspacing
A.~Rathi, D.~Sharma, and R.~K. Thapa, ``{Effects of complex-descending versus
  traditional resistance training on physical fitness abilities of female team
  sports athletes},'' \emph{Biomedical Human Kinetics}, vol.~15, no.~1, pp.
  148--158, 2023. [Online]. Available:
  \url{https://doi.org/10.2478/bhk-2023-0018}
\BIBentrySTDinterwordspacing

\bibitem{Thapa2023}
\BIBentryALTinterwordspacing
R.~K. Thapa, G.~Kumar, A.~Weldon, J.~Moran, H.~Chaabene, and
  R.~Ramirez-Campillo, ``Effects of complex-contrast training on physical
  fitness in male field hockey athletes,'' \emph{Biomedical Human Kinetics},
  vol.~15, no.~1, pp. 201--210, 2023. [Online]. Available:
  \url{https://doi.org/10.2478/bhk-2023-0024}
\BIBentrySTDinterwordspacing

\bibitem{Adamczyk2023}
\BIBentryALTinterwordspacing
J.~G. Adamczyk, ``{Support Your Recovery Needs (SYRN) - a systemic approach to
  improve sport performance},'' \emph{Biomedical Human Kinetics}, vol.~15,
  no.~1, pp. 269--279, 2023. [Online]. Available:
  \url{https://doi.org/10.2478/bhk-2023-0033}
\BIBentrySTDinterwordspacing

\bibitem{Boguszewski2024}
\BIBentryALTinterwordspacing
D.~Boguszewski, A.~Krawczyk, M.~Dębek, and J.~G. Adamczyk, ``The effects of
  foam rolling applied to delayed-onset muscle soreness of the quadriceps
  femoris after tabata training,'' \emph{Biomedical Human Kinetics}, vol.~16,
  no.~1, pp. 203--209, 2024. [Online]. Available:
  \url{https://doi.org/10.2478/bhk-2024-0021}
\BIBentrySTDinterwordspacing

\bibitem{URLBlizzardS2ClientProto}
\BIBentryALTinterwordspacing
Blizzard, ``{s2client-proto},'' 2017, acessed: 2024-09-20. [Online]. Available:
  \url{https://github.com/Blizzard/s2client-proto}
\BIBentrySTDinterwordspacing

\bibitem{URLS2Prot2017}
A.~Belicza, ``s2prot,'' \url{https://github.com/icza/s2prot}, 2016, acessed:
  2024-09-20.

\bibitem{URLBoxcars2016}
N.~Babcock, ``boxcars,'' \url{https://github.com/pnxenopoulos/awpy}, 2016,
  acessed: 2024-09-20.

\bibitem{AWPYXeno2020}
P.~Xenopoulos, ``awpy,'' \url{https://github.com/pnxenopoulos/awpy}, 2020,
  acessed: 2024-09-20.

\bibitem{Peng2020}
\BIBentryALTinterwordspacing
Q.~Peng, G.~Dickson, N.~Scelles, J.~Grix, and P.~M. Brannagan, ``Esports
  governance: Exploring stakeholder dynamics,'' \emph{Sustainability}, vol.~12,
  no.~19, 2020. [Online]. Available:
  \url{https://www.mdpi.com/2071-1050/12/19/8270}
\BIBentrySTDinterwordspacing

\bibitem{Scholz2020}
\BIBentryALTinterwordspacing
T.~M. Scholz, ``Deciphering the world of esports,'' \emph{International Journal
  on Media Management}, vol.~22, no.~1, pp. 1--12, 2020. [Online]. Available:
  \url{https://doi.org/10.1080/14241277.2020.1757808}
\BIBentrySTDinterwordspacing

\bibitem{Scholz2019}
\BIBentryALTinterwordspacing
------, \emph{{Stakeholders in the eSports Industry}}.\hskip 1em plus 0.5em
  minus 0.4em\relax Cham: Springer International Publishing, 2019, pp. 43--99.
  [Online]. Available: \url{https://doi.org/10.1007/978-3-030-11199-1_3}
\BIBentrySTDinterwordspacing

\bibitem{Claypool2017}
\BIBentryALTinterwordspacing
M.~Claypool, A.~Kica, A.~La~Manna, L.~O'Donnell, and T.~Paolillo, ``On the
  impact of software patching on gameplay for the league of legends computer
  game,'' \emph{The Computer Games Journal}, vol.~6, no.~1, pp. 33--61, 06
  2017. [Online]. Available: \url{https://doi.org/10.1007/s40869-017-0032-9}
\BIBentrySTDinterwordspacing

\bibitem{Wang2020}
\BIBentryALTinterwordspacing
Q.~Wang, Y.~Yang, Z.~Li, N.~Liu, and X.~Zhang, ``{Research on the influence of
  balance patch on players' character preference},'' \emph{Internet Research},
  vol.~30, no.~3, pp. 995--1018, 01 2020. [Online]. Available:
  \url{https://doi.org/10.1108/INTR-04-2019-0148}
\BIBentrySTDinterwordspacing

\bibitem{Zhong2022}
\BIBentryALTinterwordspacing
X.~Zhong and J.~Xu, ``Measuring the effect of game updates on player
  engagement: A cue from dota2,'' \emph{Entertainment Computing}, vol.~43, p.
  100506, 2022. [Online]. Available:
  \url{https://doi.org/10.1016/j.entcom.2022.100506}
\BIBentrySTDinterwordspacing

\bibitem{Litstudy2022}
\BIBentryALTinterwordspacing
S.~Heldens, A.~Sclocco, H.~Dreuning, B.~{van Werkhoven}, P.~Hijma, J.~Maassen,
  and R.~V. {van Nieuwpoort}, ``{litstudy: A Python package for literature
  reviews},'' \emph{SoftwareX}, vol.~20, p. 101207, 2022. [Online]. Available:
  \url{https://doi.org/10.1016/j.softx.2022.101207}
\BIBentrySTDinterwordspacing

\bibitem{Scott2022HigherED}
\BIBentryALTinterwordspacing
M.~J. Scott, R.~Summerley, N.~Besombes, C.~Connolly, J.~Gawrysiak, T.~Halevi,
  S.~E. Jenny, M.~Miljanovic, M.~Stange, T.~Taipalus, and J.~P. Williams,
  ``{Foundations for Esports Curricula in Higher Education},'' in
  \emph{Proceedings of the 2021 Working Group Reports on Innovation and
  Technology in Computer Science Education}, ser. ITiCSE-WGR '21.\hskip 1em
  plus 0.5em minus 0.4em\relax New York, NY, USA: Association for Computing
  Machinery, 2022, pp. 27--55. [Online]. Available:
  \url{https://doi.org/10.1145/3502870.3506566}
\BIBentrySTDinterwordspacing

\bibitem{Ibda2023SCHOOL}
\BIBentryALTinterwordspacing
H.~Ibda, M.~F.~A. Hakim, K.~Saifuddin, Z.~Khaq, and A.~Sunoko,
  ``\BIBforeignlanguage{English}{{Esports Games in Elementary School: A
  Systematic Literature Review}},''
  \emph{\BIBforeignlanguage{English}{International Journal on Informatics
  Visualization}}, vol.~7, no.~2, pp. 319--329, 2023. [Online]. Available:
  \url{https://doi.org/10.30630/joiv.7.2.1031}
\BIBentrySTDinterwordspacing

\bibitem{Fiskaali2020HighSchool}
\BIBentryALTinterwordspacing
A.~Fiskaali, A.~Lieberoth, and H.~Spindler,
  ``\BIBforeignlanguage{English}{{Exploring institutionalised esport in high
  school: A mixed methods study of well-being}},'' F.~P., Ed., vol.
  2020September.\hskip 1em plus 0.5em minus 0.4em\relax Dechema e.V., 2020,
  Conference paper, pp. 160--167. [Online]. Available:
  \url{https://doi.org/10.34190/GBL.20.045}
\BIBentrySTDinterwordspacing

\bibitem{Kari2016}
\BIBentryALTinterwordspacing
T.~Kari and V.-M. Karhulahti, ``{Do E-Athletes Move? A Study on Training and
  Physical Exercise in Elite E-Sports},'' \emph{International Journal of Gaming
  and Computer-Mediated Simulations}, vol.~8, no.~4, pp. 53--66, 10 2016.
  [Online]. Available: \url{https://doi.org/10.4018/IJGCMS.2016100104}
\BIBentrySTDinterwordspacing

\bibitem{Bayraktar2020}
\BIBentryALTinterwordspacing
A.~Bayrakdar, Y.~Yildiz, and I.~Bayraktar, ``{Do e-athletes move? A study on
  physical activity level and body composition in elite e-sports},''
  \emph{Physical Education of Students}, vol.~24, no.~5, p. 259i264, 10 2020.
  [Online]. Available: \url{https://doi.org/10.15561/20755279.2020.0501}
\BIBentrySTDinterwordspacing

\bibitem{Ersin2022}
\BIBentryALTinterwordspacing
A.~Ersin, H.~C. Tezeren, N.~O. Pekyavas, B.~Asal, A.~Atabey, A.~Diri, and
  I.~Gonen, ``{The Relationship Between Reaction Time and Gaming Time in
  e-sports Players},'' \emph{KINESIOLOGY}, vol.~54, no.~1, pp. 36--42, 06 2022.
  [Online]. Available: \url{https://doi.org/10.26582/k.54.1.4}
\BIBentrySTDinterwordspacing

\bibitem{Sainz2020}
\BIBentryALTinterwordspacing
I.~Sainz, D.~Collado-Mateo, and J.~D. Coso,
  ``\BIBforeignlanguage{English}{{Effect of acute caffeine intake on hit
  accuracy and reaction time in professional e-sports players}},''
  \emph{\BIBforeignlanguage{English}{Physiology and Behavior}}, vol. 224, 2020.
  [Online]. Available: \url{https://doi.org/10.1016/j.physbeh.2020.113031}
\BIBentrySTDinterwordspacing

\bibitem{Wu2024}
\BIBentryALTinterwordspacing
S.-H. Wu, Y.-C. Chen, C.-H. Chen, H.-S. Liu, Z.-X. Liu, and C.-H. Chiu,
  ``\BIBforeignlanguage{English}{{Caffeine supplementation improves the
  cognitive abilities and shooting performance of elite e-sports players: a
  crossover trial}},'' \emph{\BIBforeignlanguage{English}{Scientific Reports}},
  vol.~14, no.~1, 2024, cited by: 0; All Open Access, Gold Open Access.
  [Online]. Available: \url{https://doi.org/10.1038/s41598-024-52599-y}
\BIBentrySTDinterwordspacing

\bibitem{Qian2021}
\BIBentryALTinterwordspacing
F.~Qian and W.~Wu, ``{Simulation Training of E-Sports Players Based on Wireless
  Sensor Network},'' \emph{Wireless Communications and Mobile Computing}, vol.
  2021, 01 2021. [Online]. Available:
  \url{https://doi.org/10.1155/2021/9636951}
\BIBentrySTDinterwordspacing

\bibitem{Argiles2022}
\BIBentryALTinterwordspacing
M.~Argiles, L.~Quevedo-Junyent, and G.~Erickson, ``{Topical Review: Optometric
  Considerations in Sports Versus E-Sports},'' \emph{PERCEPTUAL AND MOTOR
  SKILLS}, vol. 129, no.~3, pp. 731--746, 06 2022. [Online]. Available:
  \url{https://doi.org/10.1177/00315125211073401}
\BIBentrySTDinterwordspacing

\bibitem{Goulart2023}
\BIBentryALTinterwordspacing
J.~B. Goulart, L.~S. Aitken, S.~Siddiqui, M.~Cuevas, J.~Cardenas, K.~M.
  Beathard, and S.~E. Riechman, ``\BIBforeignlanguage{English}{{Nutrition,
  lifestyle, and cognitive performance in esport athletes}},''
  \emph{\BIBforeignlanguage{English}{Frontiers in Nutrition}}, vol.~10, 2023.
  [Online]. Available: \url{https://doi.org/10.3389/fnut.2023.1120303}
\BIBentrySTDinterwordspacing

\bibitem{Yakovlev2020}
\BIBentryALTinterwordspacing
L.~Yakovlev, N.~Syrov, N.~Görtz, and A.~Kaplan,
  ``\BIBforeignlanguage{English}{{BCI-Controlled Motor Imagery Training Can
  Improve Performance in e-Sports}},''
  \emph{\BIBforeignlanguage{English}{Communications in Computer and Information
  Science}}, vol. 1224 CCIS, pp. 581 -- 586, 2020, cited by: 9; Conference
  name: 22nd International Conference on Human-Computer Interaction, HCII 2020;
  Conference date: 19 July 2020 through 24 July 2020; Conference code: 242529.
  [Online]. Available: \url{https://doi.org/10.1007/978-3-030-50726-8_76}
\BIBentrySTDinterwordspacing

\bibitem{Moritz2023}
\BIBentryALTinterwordspacing
S.~E. Moritz, ``\BIBforeignlanguage{English}{Translating the applied model of
  deliberate imagery use to esports},''
  \emph{\BIBforeignlanguage{English}{Journal of Imagery Research in Sport and
  Physical Activity}}, vol.~18, no. 1 s, 2023, cited by: 6. [Online].
  Available: \url{https://doi.org/10.1515/jirspa-2023-0014}
\BIBentrySTDinterwordspacing

\bibitem{Cumming2023}
\BIBentryALTinterwordspacing
J.~Cumming and M.~L. Quinton, ``\BIBforeignlanguage{English}{Developing imagery
  ability in esport athletes using layered stimulus response training},''
  \emph{\BIBforeignlanguage{English}{Journal of Imagery Research in Sport and
  Physical Activity}}, vol.~18, no. 1 s, 2023, cited by: 3; All Open Access,
  Green Open Access, Hybrid Gold Open Access. [Online]. Available:
  \url{https://doi.org/10.1515/jirspa-2022-0024}
\BIBentrySTDinterwordspacing

\bibitem{SanzMatesanz2024}
\BIBentryALTinterwordspacing
M.~Sanz-Matesanz, L.~M. Martínez-Aranda, and G.~M. Gea-García,
  ``\BIBforeignlanguage{English}{{Effects of a Physical Training Program on
  Cognitive and Physical Performance and Health-Related Variables in
  Professional esports Players: A Pilot Study}},''
  \emph{\BIBforeignlanguage{English}{Applied Sciences (Switzerland)}}, vol.~14,
  no.~7, 2024, cited by: 1; All Open Access, Gold Open Access. [Online].
  Available: \url{https://doi.org/10.3390/app14072845}
\BIBentrySTDinterwordspacing

\bibitem{Szepne2019physical}
\BIBentryALTinterwordspacing
H.~V. Szépné, L.~Csernoch, and I.~Balatoni,
  ``\BIBforeignlanguage{English}{{E-sports versus physical activity among
  adolescents}},'' \emph{\BIBforeignlanguage{English}{Baltic Journal of Health
  and Physical Activity}}, vol.~11, no.~6, pp. 38 -- 47, 2019, cited by: 8; All
  Open Access, Gold Open Access. [Online]. Available:
  \url{http://dx.doi.org/10.29359/BJHPA.2019.Suppl.2.06}
\BIBentrySTDinterwordspacing

\bibitem{Fletcher2021}
\BIBentryALTinterwordspacing
B.~Fletcher and D.~James, ``{Grassroots Esports Players: Improving Esports
  Cognitive Skills Through Incentivising Physical Exercise},'' in \emph{Serious
  Games: Joint International Conference, JCSG 2021, Virtual Event, January
  12-13, 2022, Proceedings}.\hskip 1em plus 0.5em minus 0.4em\relax Berlin,
  Heidelberg: Springer-Verlag, 2021, pp. 200--212. [Online]. Available:
  \url{https://doi.org/10.1007/978-3-030-88272-3_15}
\BIBentrySTDinterwordspacing

\bibitem{Difrancisco-Donoghue2021}
\BIBentryALTinterwordspacing
J.~Difrancisco-Donoghue, S.~E. Jenny, P.~C. Douris, S.~Ahmad, K.~Yuen,
  T.~Hassan, H.~Gan, K.~Abraham, and A.~Sousa,
  ``\BIBforeignlanguage{English}{{Breaking up prolonged sitting with a 6 min
  walk improves executive function in women and men esports players: A
  randomised trial}},'' \emph{\BIBforeignlanguage{English}{BMJ Open Sport and
  Exercise Medicine}}, vol.~7, no.~3, 2021, cited by: 7; All Open Access, Gold
  Open Access. [Online]. Available:
  \url{https://doi.org/10.1136/bmjsem-2021-001118}
\BIBentrySTDinterwordspacing

\bibitem{Ekdahl2022}
\BIBentryALTinterwordspacing
D.~Ekdahl, ``\BIBforeignlanguage{English}{Both physical and virtual: On
  immediacy in esports},'' \emph{\BIBforeignlanguage{English}{Frontiers in
  Sports and Active Living}}, vol.~4, 2022, cited by: 12; All Open Access, Gold
  Open Access, Green Open Access. [Online]. Available:
  \url{https://doi.org/10.3389/fspor.2022.883765}
\BIBentrySTDinterwordspacing

\bibitem{Pereira2019}
\BIBentryALTinterwordspacing
A.~M. Pereira, J.~Brito, P.~Figueiredo, and E.~Verhagen,
  ``\BIBforeignlanguage{English}{{Virtual sports deserve real sports medical
  attention}},'' \emph{\BIBforeignlanguage{English}{BMJ Open Sport and Exercise
  Medicine}}, vol.~5, no.~1, 2019, cited by: 34; All Open Access, Gold Open
  Access, Green Open Access. [Online]. Available:
  \url{https://doi.org/10.1136/bmjsem-2019-000606}
\BIBentrySTDinterwordspacing

\bibitem{Law2023224}
\BIBentryALTinterwordspacing
A.~Law, G.~Ho, and M.~Moore, ``\BIBforeignlanguage{English}{{Care of the
  Esports Athlete}},'' \emph{\BIBforeignlanguage{English}{Current Sports
  Medicine Reports}}, vol.~22, no.~6, pp. 224 -- 229, 2023, cited by: 0; All
  Open Access, Bronze Open Access. [Online]. Available:
  \url{https://doi.org/10.1249/jsr.0000000000001077}
\BIBentrySTDinterwordspacing

\bibitem{McGee2021415}
\BIBentryALTinterwordspacing
C.~McGee, M.~Hwu, L.~L. Nicholson, and K.~K. Ho,
  ``\BIBforeignlanguage{English}{{More than a game: Musculoskeletal injuries
  and a key role for the physical therapist in esports}},''
  \emph{\BIBforeignlanguage{English}{Journal of Orthopaedic and Sports Physical
  Therapy}}, vol.~51, no.~9, pp. 415 -- 417, 2021, cited by: 6. [Online].
  Available: \url{https://doi.org/10.2519/jospt.2021.0109}
\BIBentrySTDinterwordspacing

\bibitem{Migliore2021}
\BIBentryALTinterwordspacing
L.~Migliore, \emph{{Prevention of Esports Injuries}}.\hskip 1em plus 0.5em
  minus 0.4em\relax Cham: Springer International Publishing, 2021, pp.
  213--240. [Online]. Available:
  \url{https://doi.org/10.1007/978-3-030-73610-1_9}
\BIBentrySTDinterwordspacing

\bibitem{Lindberg2020Pain}
\BIBentryALTinterwordspacing
L.~Lindberg, S.~B. Nielsen, M.~Damgaard, O.~R. Sloth, M.~S. Rathleff, and C.~L.
  Straszek, ``\BIBforeignlanguage{English}{{Musculoskeletal pain is common in
  competitive gaming: A cross-sectional study among Danish esports
  athletes}},'' \emph{\BIBforeignlanguage{English}{BMJ Open Sport and Exercise
  Medicine}}, vol.~6, no.~1, 2020, cited by: 26; All Open Access, Gold Open
  Access, Green Open Access. [Online]. Available:
  \url{https://doi.org/10.1136/bmjsem-2020-000799}
\BIBentrySTDinterwordspacing

\bibitem{Manci2024}
\BIBentryALTinterwordspacing
E.~Manci, U.~Gencturk, E.~Günay, C.~Guducu, F.~Herold, and C.~S. Bediz,
  ``\BIBforeignlanguage{English}{{The influence of acute sprint exercise on
  cognition, gaming performance, and cortical hemodynamics in esports players
  and age-matched controls}},'' \emph{\BIBforeignlanguage{English}{Current
  Psychology}}, 2024, cited by: 0; All Open Access, Hybrid Gold Open Access.
  [Online]. Available: \url{https://doi.org/10.1007/s12144-024-05750-x}
\BIBentrySTDinterwordspacing

\bibitem{Lam2020Eathletes}
\BIBentryALTinterwordspacing
A.~T.~W. Lam, T.~P. Perera, K.~B.~A. Quirante, A.~Wilks, A.~J. Ionas, and G.~D.
  Baxter, ``\BIBforeignlanguage{English}{{E-athletes' lifestyle behaviors,
  physical activity habits, and overall health and wellbeing: a systematic
  review}},'' \emph{\BIBforeignlanguage{English}{Physical Therapy Reviews}},
  vol.~25, no. 5-6, pp. 449 -- 461, 2020, cited by: 8. [Online]. Available:
  \url{https://www.scopus.com/inward/record.uri?eid=2-s2.0-85097938438&doi=10.1080%2f10833196.2020.1843352&partnerID=40&md5=b34fffd03736ef66d48f82927c3953ef}
\BIBentrySTDinterwordspacing

\bibitem{Voisin202232}
\BIBentryALTinterwordspacing
N.~Voisin, N.~Besombes, and S.~Laffage-Cosnier,
  ``\BIBforeignlanguage{English}{{Are Esports Players Inactive? A Systematic
  Review}},'' \emph{\BIBforeignlanguage{English}{Physical Culture and Sport,
  Studies and Research}}, vol.~97, no.~1, pp. 32 -- 52, 2022, cited by: 0; All
  Open Access, Gold Open Access, Green Open Access. [Online]. Available:
  \url{https://doi.org/10.2478/pcssr-2022-0022}
\BIBentrySTDinterwordspacing

\bibitem{Nicholson2024}
\BIBentryALTinterwordspacing
M.~Nicholson, C.~Thompson, D.~Poulus, T.~Pavey, R.~Robergs, V.~Kelly, and
  C.~McNulty, ``\BIBforeignlanguage{English}{{Physical Activity and
  Self-Determination towards Exercise among Esports Athletes}},''
  \emph{\BIBforeignlanguage{English}{Sports Medicine - Open}}, vol.~10, no.~1,
  2024, cited by: 0; All Open Access, Gold Open Access. [Online]. Available:
  \url{https://doi.org/10.1186/s40798-024-00700-0}
\BIBentrySTDinterwordspacing

\bibitem{Ketelhut2021}
\BIBentryALTinterwordspacing
S.~Ketelhut, A.~L. Martin-Niedecken, P.~Zimmermann, and C.~R. Nigg,
  ``\BIBforeignlanguage{English}{{Physical Activity and Health Promotion in
  Esports and Gaming-Discussing Unique Opportunities for an Unprecedented
  Cultural Phenomenon}},'' \emph{\BIBforeignlanguage{English}{Frontiers in
  Sports and Active Living}}, vol.~3, 2021, cited by: 19; All Open Access, Gold
  Open Access. [Online]. Available:
  \url{https://doi.org/10.3389/fspor.2021.693700}
\BIBentrySTDinterwordspacing

\bibitem{Trocchio2021}
\BIBentryALTinterwordspacing
L.~Trocchio, \emph{\BIBforeignlanguage{English}{{Nutrition for the video
  gamer}}}.\hskip 1em plus 0.5em minus 0.4em\relax Springer International
  Publishing, 2021, cited by: 0. [Online]. Available:
  \url{https://doi.org/10.1007/978-3-030-73610-1_6}
\BIBentrySTDinterwordspacing

\bibitem{Nagorsky2020}
\BIBentryALTinterwordspacing
E.~Nagorsky and J.~Wiemeyer, ``{The structure of performance and training in
  esports},'' \emph{PLOS ONE}, vol.~15, no.~8, AUG 25 2020. [Online].
  Available: \url{https://doi.org/10.1371/journal.pone.0237584}
\BIBentrySTDinterwordspacing

\bibitem{Bialecki2022}
\BIBentryALTinterwordspacing
A.~Białecki, R.~Białecki, and J.~Gajewski, ``{Redefining Sports: Esports,
  Environments, Signals and Functions},'' \emph{International Journal of
  Electronics and Telecommunications}, vol. vol. 68, no. No 3, pp. 541--548,
  2022. [Online]. Available: \url{http://dx.doi.org/10.24425/ijet.2022.141272}
\BIBentrySTDinterwordspacing

\bibitem{Bikas2023}
\BIBentryALTinterwordspacing
I.~Bikas, J.~Pfau, T.~Muender, D.~Alexandrovsky, and R.~Malaka, ``{Grinding to
  a Halt: The Effects of Long Play Sessions on Player Performance in Video
  Games},'' in \emph{Companion Proceedings of the Annual Symposium on
  Computer-Human Interaction in Play}, ser. CHI PLAY Companion '23.\hskip 1em
  plus 0.5em minus 0.4em\relax New York, NY, USA: Association for Computing
  Machinery, 2023, pp. 36--42. [Online]. Available:
  \url{https://doi.org/10.1145/3573382.3616073}
\BIBentrySTDinterwordspacing

\bibitem{He2021Updates}
\BIBentryALTinterwordspacing
Y.~He, C.~Tran, J.~Jiang, K.~Burghardt, E.~Ferrara, E.~Zheleva, and K.~Lerman,
  ``{Heterogeneous Effects of Software Patches in aMultiplayer Online Battle
  Arena Game},'' F.~A., P.~J., C.~A.A., A.~A.A., and H.~C., Eds.\hskip 1em plus
  0.5em minus 0.4em\relax Association for Computing Machinery, 2021, Conference
  paper. [Online]. Available: \url{https://doi.org/10.1145/3472538.3472550}
\BIBentrySTDinterwordspacing

\bibitem{Yu2021Updates}
\BIBentryALTinterwordspacing
Y.~Yu, B.-H. Nguyen, F.~Yu, and V.-N. Huynh, ``{Esports Game Updates and Player
  Perception: Data Analysis of PUBG Steam Reviews},'' S.~T., B.~L.T., M.~Y.,
  R.~T., H.~K., and Y.~T., Eds., vol. 2021-November.\hskip 1em plus 0.5em minus
  0.4em\relax Institute of Electrical and Electronics Engineers Inc., 2021.
  [Online]. Available: \url{https://doi.org/10.1109/KSE53942.2021.9648670}
\BIBentrySTDinterwordspacing

\bibitem{Białecki2023SC2EGSetDataset}
\BIBentryALTinterwordspacing
A.~Białecki, N.~Jakubowska, P.~Dobrowolski, P.~Białecki, L.~Krupiński,
  A.~Szczap, R.~Białecki, and J.~Gajewski,
  ``\BIBforeignlanguage{English}{{SC2EGSet: StarCraft II Esport Replay and
  Game-state Dataset}},'' \emph{\BIBforeignlanguage{English}{Scientific Data}},
  vol.~10, no.~1, 2023. [Online]. Available:
  \url{https://www.scopus.com/inward/record.uri?eid=2-s2.0-85170344767&doi=10.1038%2fs41597-023-02510-7&partnerID=40&md5=0080b5862eb0aa49006c89ff225fc291}
\BIBentrySTDinterwordspacing

\bibitem{Wu2023Dataset}
\BIBentryALTinterwordspacing
H.~Wu, Y.~Zong, J.~Zhang, and K.~Huang, ``{MSC: A Dataset for Macro-Management
  in StarCraft II},'' 2023. [Online]. Available:
  \url{https://doi.org/10.48550/arXiv.1710.03131}
\BIBentrySTDinterwordspacing

\bibitem{Gao2022Dataset}
\BIBentryALTinterwordspacing
Y.~Gao, ``{PGD: A Large-scale Professional Go Dataset for Data-driven
  Analytics},'' in \emph{2022 IEEE Conference on Games (CoG)}.\hskip 1em plus
  0.5em minus 0.4em\relax IEEE Press, 2022, pp. 284--291. [Online]. Available:
  \url{https://doi.org/10.1109/CoG51982.2022.9893704}
\BIBentrySTDinterwordspacing

\bibitem{Tanaka2021Dataset}
\BIBentryALTinterwordspacing
T.~Tanaka and E.~Simo-Serra, ``\BIBforeignlanguage{English}{{LoL-V2T:
  Large-scale esports video description dataset}}.''\hskip 1em plus 0.5em minus
  0.4em\relax IEEE Computer Society, 2021, Conference paper, pp. 4552--4561.
  [Online]. Available:
  \url{https://www.scopus.com/inward/record.uri?eid=2-s2.0-85113598362&doi=10.1109%2fCVPRW53098.2021.00513&partnerID=40&md5=d2d10da41b796074679059aee8bae4e1}
\BIBentrySTDinterwordspacing

\bibitem{Xu2023Dataset}
\BIBentryALTinterwordspacing
J.~H. Xu, Y.~Nakano, L.~Kong, and K.~Iizuka, ``{CS-lol: a Dataset of Viewer
  Comment with Scene in E-sports Live-streaming},'' in \emph{Proceedings of the
  2023 Conference on Human Information Interaction and Retrieval}, ser. CHIIR
  '23.\hskip 1em plus 0.5em minus 0.4em\relax New York, NY, USA: Association
  for Computing Machinery, 2023, pp. 422--426. [Online]. Available:
  \url{https://doi.org/10.1145/3576840.3578334}
\BIBentrySTDinterwordspacing

\bibitem{Xenopoulos2022dataset}
\BIBentryALTinterwordspacing
P.~Xenopoulos and C.~Silva, ``{ESTA: An Esports Trajectory and Action
  Dataset},'' 2022. [Online]. Available:
  \url{https://doi.org/10.48550/arXiv.2209.09861}
\BIBentrySTDinterwordspacing

\bibitem{Horst2021}
\BIBentryALTinterwordspacing
R.~Horst, S.~M. Zander, and R.~D\"{o}rner, ``{CS:Show - An Interactive Visual
  Analysis Tool for First-Person Shooter eSports Match Data},'' in
  \emph{Entertainment Computing - ICEC 2021: 20th IFIP TC 14 International
  Conference, ICEC 2021, Coimbra, Portugal, November 2-5, 2021,
  Proceedings}.\hskip 1em plus 0.5em minus 0.4em\relax Berlin, Heidelberg:
  Springer-Verlag, 2021, pp. 15--27. [Online]. Available:
  \url{https://doi.org/10.1007/978-3-030-89394-1_2}
\BIBentrySTDinterwordspacing

\bibitem{Afonso201933069}
\BIBentryALTinterwordspacing
A.~P. Afonso, M.~B. Carmo, T.~Gonçalves, and P.~Vieira,
  ``\BIBforeignlanguage{English}{{VisuaLeague: Player performance analysis
  using spatial-temporal data}},''
  \emph{\BIBforeignlanguage{English}{Multimedia Tools and Applications}},
  vol.~78, no.~23, pp. 33\,069 -- 33\,090, 2019, cited by: 16. [Online].
  Available: \url{https://doi.org/10.1007/s11042-019-07952-z}
\BIBentrySTDinterwordspacing

\bibitem{Kim202134189}
\BIBentryALTinterwordspacing
S.~Kim, D.~Kim, H.~Ahn, and B.~Ahn,
  ``\BIBforeignlanguage{English}{{Implementation of user playstyle coaching
  using video processing and statistical methods in league of legends}},''
  \emph{\BIBforeignlanguage{English}{Multimedia Tools and Applications}},
  vol.~80, no. 26-27, p. 34189 – 34201, 2021, cited by: 2. [Online].
  Available:
  \url{https://www.scopus.com/inward/record.uri?eid=2-s2.0-85089786345&doi=10.1007%2fs11042-020-09413-4&partnerID=40&md5=a70148b78a93f59285f1f6de4fbb9734}
\BIBentrySTDinterwordspacing

\bibitem{Spiricheva2019}
\BIBentryALTinterwordspacing
N.~Spiricheva and S.~Akulich, ``\BIBforeignlanguage{English}{{Infographics
  Application for Visualization of E-Sports Gaming Activity}}.''\hskip 1em plus
  0.5em minus 0.4em\relax Institute of Electrical and Electronics Engineers
  Inc., 2019, Conference paper. [Online]. Available:
  \url{https://doi.org/10.1109/ICISCT47635.2019.9011849}
\BIBentrySTDinterwordspacing

\bibitem{Broek2019}
\BIBentryALTinterwordspacing
W.~van~den Broek, G.~Wallner, and R.~Bernhaupt, ``{Modata -- Improving Dota 2
  Experience and Spectatorship through Tangible Gameplay Visualization},'' in
  \emph{Extended Abstracts of the Annual Symposium on Computer-Human
  Interaction in Play Companion Extended Abstracts}, ser. CHI PLAY '19 Extended
  Abstracts.\hskip 1em plus 0.5em minus 0.4em\relax New York, NY, USA:
  Association for Computing Machinery, 2019, pp. 723--730. [Online]. Available:
  \url{https://doi.org/10.1145/3341215.3356284}
\BIBentrySTDinterwordspacing

\bibitem{Font2019285}
\BIBentryALTinterwordspacing
J.~M. Font and T.~Mahlmann, ``\BIBforeignlanguage{English}{{DOTA 2 bot
  competition}},'' \emph{\BIBforeignlanguage{English}{IEEE Transactions on
  Games}}, vol.~11, no.~3, pp. 285--289, 2019, cited by: 16; All Open Access,
  Green Open Access. [Online]. Available:
  \url{https://doi.org/10.1109/TG.2018.2834566}
\BIBentrySTDinterwordspacing

\bibitem{Goodway2015}
\BIBentryALTinterwordspacing
J.~D. Goodway and L.~E. Robinson, ``{Developmental Trajectories in Early Sport
  Specialization: A Case for Early Sampling from a Physical Growth and Motor
  Development Perspective},'' \emph{Kinesiology Review}, vol.~4, no.~3, pp.
  267--278, 2015. [Online]. Available:
  \url{https://doi.org/10.1123/kr.2015-0028}
\BIBentrySTDinterwordspacing

\bibitem{Szymanska2016}
\BIBentryALTinterwordspacing
E.~Szymańska, S.~Żak, E.~Mleczko, R.~Nieroda, and T.~Klocek, ``{ONCE MORE ON
  THE METHODS FOR DETERMINING SENSITIVE AND CRITICAL PERIODS IN THE MOTOR
  DEVELOPMENT OF CHILDREN AND THE YOUTH},'' \emph{Journal of Kinesiology and
  Exercise Sciences}, vol.~26, pp. 29--50, 10 2016. [Online]. Available:
  \url{http://dx.doi.org/10.5604/01.3001.0010.0926}
\BIBentrySTDinterwordspacing

\bibitem{Arslan2024Motivation}
\BIBentryALTinterwordspacing
Y.~Arslan, C.~Suveren, and T.~A. Durhan, ``{E-Sports Participation Motivation
  from the Perspective of Sports Sciences Students},'' \emph{ANNALS OF APPLIED
  SPORT SCIENCE}, vol.~12, no.~1, 2024. [Online]. Available:
  \url{http://dx.doi.org/10.61186/aassjournal.1329}
\BIBentrySTDinterwordspacing

\bibitem{Jasny2019}
M.~Jasny, \emph{{Sportowy wymiar maniaczenia przy komputerze, czyli
  kształtowanie sprawności fizycznej w ramach treningu w e-sporcie}}, 09
  2019, pp. 57--70.

\bibitem{Anders2016PEAK}
A.~Ericsson and R.~Pool, \emph{{Peak: Secrets from the New Science of
  Expertise}}.\hskip 1em plus 0.5em minus 0.4em\relax Mariner Books, 2016.

\bibitem{Brenner2016}
\BIBentryALTinterwordspacing
J.~S. Brenner, C.~O.~S. MEDICINE, and FITNESS, ``{Sports Specialization and
  Intensive Training in Young Athletes},'' \emph{Pediatrics}, vol. 138, no.~3,
  p. e20162148, 09 2016. [Online]. Available:
  \url{https://doi.org/10.1542/peds.2016-2148}
\BIBentrySTDinterwordspacing

\bibitem{Myer2016}
\BIBentryALTinterwordspacing
G.~D. Myer, N.~Jayanthi, J.~P. DiFiori, A.~D. Faigenbaum, A.~W. Kiefer,
  D.~Logerstedt, and L.~J. Micheli, ``{Sports Specialization, Part II:
  Alternative Solutions to Early Sport Specialization in Youth Athletes},''
  \emph{Sports Health}, vol.~8, no.~1, pp. 65--73, 2016, pMID: 26517937.
  [Online]. Available: \url{https://doi.org/10.1177/1941738115614811}
\BIBentrySTDinterwordspacing

\bibitem{Myer2015}
\BIBentryALTinterwordspacing
G.~D. Myer, N.~Jayanthi, J.~P. Difiori, A.~D. Faigenbaum, A.~W. Kiefer,
  D.~Logerstedt, and L.~J. Micheli, ``{Sport Specialization, Part I: Does Early
  Sports Specialization Increase Negative Outcomes and Reduce the Opportunity
  for Success in Young Athletes?}'' \emph{Sports Health}, vol.~7, no.~5, pp.
  437--442, 2015, pMID: 26502420. [Online]. Available:
  \url{https://doi.org/10.1177/1941738115598747}
\BIBentrySTDinterwordspacing

\bibitem{Granacher2017}
\BIBentryALTinterwordspacing
U.~Granacher and R.~Borde, ``{Effects of Sport-Specific Training during the
  Early Stages of Long-Term Athlete Development on Physical Fitness, Body
  Composition, Cognitive, and Academic Performances},'' \emph{Frontiers in
  Physiology}, vol.~8, 2017. [Online]. Available:
  \url{https://doi.org/10.3389/fphys.2017.00810}
\BIBentrySTDinterwordspacing

\bibitem{Balyi2001}
I.~Balyi, ``{Sport System Building and Long-term athlete development in
  Britisch Columbia, Canada},'' \emph{Sports Medicine BC}, 2001.

\bibitem{BalyiLTAD2013}
\BIBentryALTinterwordspacing
I.~Balyi, R.~Way, and C.~Higgs, \emph{{Long-Term Athlete Development}}, 01
  2013. [Online]. Available: \url{http://dx.doi.org/10.5040/9781492596318}
\BIBentrySTDinterwordspacing

\bibitem{Meredith2015}
\BIBentryALTinterwordspacing
R.~Meredith, ``{Sensitive and critical periods during neurotypical and aberrant
  neurodevelopment: A framework for neurodevelopmental disorders},''
  \emph{{Neuroscience \& Biobehavioral Reviews}}, vol.~50, pp. 180--188, 2015,
  brain, Memory and Development: The Imprint of Gabriel Horn. [Online].
  Available: \url{https://doi.org/10.1016/j.neubiorev.2014.12.001}
\BIBentrySTDinterwordspacing

\bibitem{Fuhrmann2015}
\BIBentryALTinterwordspacing
D.~Fuhrmann, L.~J. Knoll, and S.-J. Blakemore, ``{Adolescence as a Sensitive
  Period of Brain Development},'' \emph{Trends in Cognitive Sciences}, vol.~19,
  no.~10, pp. 558--566, 10 2015. [Online]. Available:
  \url{https://doi.org/10.1016/j.tics.2015.07.008}
\BIBentrySTDinterwordspacing

\bibitem{Sawan2020MRAR}
\BIBentryALTinterwordspacing
N.~Sawan, A.~Eltweri, C.~De~Lucia, L.~P.~L. Cavaliere, A.~Faccia, and N.~R.
  Moşteanu, ``\BIBforeignlanguage{English}{{Mixed and Augmented Reality
  Applications in the Sport Industry}}.''\hskip 1em plus 0.5em minus
  0.4em\relax Association for Computing Machinery, 2020, Conference paper, pp.
  55--59. [Online]. Available: \url{https://doi.org/10.1145/3446922.3446932}
\BIBentrySTDinterwordspacing

\bibitem{Masasuke2023ARCHERY}
\BIBentryALTinterwordspacing
M.~Yasumoto, ``{Bow Device for Accurate Reproduction of Archery in xR
  Environment},'' in \emph{Virtual, Augmented and Mixed Reality: 15th
  International Conference, VAMR 2023, Held as Part of the 25th HCI
  International Conference, HCII 2023, Copenhagen, Denmark, July 23–28, 2023,
  Proceedings}.\hskip 1em plus 0.5em minus 0.4em\relax Berlin, Heidelberg:
  Springer-Verlag, 2023, pp. 203--214. [Online]. Available:
  \url{https://doi.org/10.1007/978-3-031-35634-6_15}
\BIBentrySTDinterwordspacing

\bibitem{Masasuke2022ARCHERY}
\BIBentryALTinterwordspacing
------, ``{Evaluation of the Difference in the Reality of the Bow Device with
  and Without Arrows},'' in \emph{Design, User Experience, and Usability:
  Design Thinking and Practice in Contemporary and Emerging Technologies: 11th
  International Conference, DUXU 2022, Held as Part of the 24th HCI
  International Conference, HCII 2022, Virtual Event, June 26 – July 1, 2022,
  Proceedings, Part III}.\hskip 1em plus 0.5em minus 0.4em\relax Berlin,
  Heidelberg: Springer-Verlag, 2022, pp. 428--441. [Online]. Available:
  \url{https://doi.org/10.1007/978-3-031-05906-3_32}
\BIBentrySTDinterwordspacing

\bibitem{Silver2016}
\BIBentryALTinterwordspacing
D.~Silver, A.~Huang, C.~J. Maddison, A.~Guez, L.~Sifre, G.~van~den Driessche,
  J.~Schrittwieser, I.~Antonoglou, V.~Panneershelvam, M.~Lanctot, S.~Dieleman,
  D.~Grewe, J.~Nham, N.~Kalchbrenner, I.~Sutskever, T.~Lillicrap, M.~Leach,
  K.~Kavukcuoglu, T.~Graepel, and D.~Hassabis, ``{Mastering the game of Go with
  deep neural networks and tree search},'' \emph{Nature}, vol. 529, no. 7587,
  pp. 484--489, 01 2016. [Online]. Available:
  \url{https://doi.org/10.1038/nature16961}
\BIBentrySTDinterwordspacing

\bibitem{Schrittwieser2020}
\BIBentryALTinterwordspacing
J.~Schrittwieser, I.~Antonoglou, T.~Hubert, K.~Simonyan, L.~Sifre, S.~Schmitt,
  A.~Guez, E.~Lockhart, D.~Hassabis, T.~Graepel, T.~Lillicrap, and D.~Silver,
  ``{Mastering Atari, Go, chess and shogi by planning with a learned model},''
  \emph{Nature}, vol. 588, no. 7839, pp. 604--609, 12 2020. [Online].
  Available: \url{https://doi.org/10.1038/s41586-020-03051-4}
\BIBentrySTDinterwordspacing

\bibitem{Silver2018}
\BIBentryALTinterwordspacing
D.~Silver, T.~Hubert, J.~Schrittwieser, I.~Antonoglou, M.~Lai, A.~Guez,
  M.~Lanctot, L.~Sifre, D.~Kumaran, T.~Graepel, T.~Lillicrap, K.~Simonyan, and
  D.~Hassabis, ``{A general reinforcement learning algorithm that masters
  chess, shogi, and Go through self-play},'' \emph{Science}, vol. 362, no.
  6419, pp. 1140--1144, 2018. [Online]. Available:
  \url{https://doi.org/10.1126/science.aar6404}
\BIBentrySTDinterwordspacing

\bibitem{Vinyals2019350}
O.~Vinyals, I.~Babuschkin, W.~M. Czarnecki, M.~Mathieu, A.~Dudzik, J.~Chung,
  D.~H. Choi, R.~Powell, T.~Ewalds, P.~Georgiev, J.~Oh, D.~Horgan, M.~Kroiss,
  I.~Danihelka, A.~Huang, L.~Sifre, T.~Cai, J.~P. Agapiou, M.~Jaderberg, A.~S.
  Vezhnevets, R.~Leblond, T.~Pohlen, V.~Dalibard, D.~Budden, Y.~Sulsky,
  J.~Molloy, T.~L. Paine, C.~Gulcehre, Z.~Wang, T.~Pfaff, Y.~Wu, R.~Ring,
  D.~Yogatama, D.~Wünsch, K.~McKinney, O.~Smith, T.~Schaul, T.~Lillicrap,
  K.~Kavukcuoglu, D.~Hassabis, C.~Apps, and D.~Silver,
  ``\BIBforeignlanguage{English}{{Grandmaster level in StarCraft II using
  multi-agent reinforcement learning}},''
  \emph{\BIBforeignlanguage{English}{Nature}}, vol. 575, no. 7782, pp.
  350--354, 2019, cited by: 2211.

\bibitem{openai2019dota2largescale}
\BIBentryALTinterwordspacing
OpenAI, :, C.~Berner, G.~Brockman, B.~Chan, V.~Cheung, P.~Dębiak, C.~Dennison,
  D.~Farhi, Q.~Fischer, S.~Hashme, C.~Hesse, R.~Józefowicz, S.~Gray,
  C.~Olsson, J.~Pachocki, M.~Petrov, H.~P. d.~O.~Pinto, J.~Raiman, T.~Salimans,
  J.~Schlatter, J.~Schneider, S.~Sidor, I.~Sutskever, J.~Tang, F.~Wolski, and
  S.~Zhang, ``{Dota 2 with Large Scale Deep Reinforcement Learning},'' 2019.
  [Online]. Available: \url{https://arxiv.org/abs/1912.06680}
\BIBentrySTDinterwordspacing

\bibitem{raiman2019longtermplanning}
\BIBentryALTinterwordspacing
J.~Raiman, S.~Zhang, and F.~Wolski, ``{Long-Term Planning and Situational
  Awareness in OpenAI Five},'' 2019. [Online]. Available:
  \url{https://arxiv.org/abs/1912.06721}
\BIBentrySTDinterwordspacing

\bibitem{Sutton2018}
R.~S. Sutton and A.~G. Barto, \emph{Reinforcement Learning: An
  Introduction}.\hskip 1em plus 0.5em minus 0.4em\relax Cambridge, MA, USA: A
  Bradford Book, 2018.

\bibitem{Damastuti2024}
F.~A. Damastuti, K.~Firmansyah, Y.~M. Arif, T.~Dutono, A.~Barakbah, and
  M.~Hariadi, ``{Dynamic Level of Difficulties Using Q-Learning and Fuzzy
  Logic},'' \emph{IEEE Access}, pp. 1--1, 2024.

\bibitem{Rosa2023}
M.~P.~C. Rosa, C.~D. Castanho, T.~B.~P. e~Silva, M.~M. Sarmet, and R.~P.
  Jacobi, ``{Dynamic Difficulty Adjustment by Performance and Player Profile in
  Platform Game},'' in \emph{Entertainment Computing -- ICEC 2023},
  P.~Ciancarini, A.~Di~Iorio, H.~Hlavacs, and F.~Poggi, Eds.\hskip 1em plus
  0.5em minus 0.4em\relax Singapore: Springer Nature Singapore, 2023, pp.
  3--16.

\bibitem{Silva2017}
\BIBentryALTinterwordspacing
M.~P. Silva, V.~do~Nascimento~Silva, and L.~Chaimowicz, ``{Dynamic difficulty
  adjustment on MOBA games},'' \emph{Entertainment Computing}, vol.~18, pp.
  103--123, 2017. [Online]. Available:
  \url{https://doi.org/10.1016/j.entcom.2016.10.002}
\BIBentrySTDinterwordspacing

\bibitem{Mortazavi2024}
\BIBentryALTinterwordspacing
F.~Mortazavi, H.~Moradi, and A.-H. Vahabie, ``Dynamic difficulty adjustment
  approaches in video games: a systematic literature review,'' \emph{Multimedia
  Tools and Applications}, Mar 2024. [Online]. Available:
  \url{https://doi.org/10.1007/s11042-024-18768-x}
\BIBentrySTDinterwordspacing

\end{thebibliography}

\end{document}